\pdfoutput=1
\documentclass[fleqn,usenatbib]{mnras}

\usepackage{newtxtext,newtxmath}

\usepackage[T1]{fontenc}

\DeclareRobustCommand{\VAN}[3]{#2}
\let\VANthebibliography\thebibliography
\def\thebibliography{\DeclareRobustCommand{\VAN}[3]{##3}\VANthebibliography}

\usepackage{graphicx}	
\usepackage{amsmath}	

\newcommand{\target}{G23.6644\texorpdfstring{$-$}{-}0.0372}
\newcommand{\nodata}{}

\title[Galactic plane variable radio sources]{
Structural and spectral properties of Galactic plane variable radio sources 
}

\author[J. Yang et al.]{Jun Yang,$^{1,2}$\thanks{E-mail: jun.yang@chalmers.se} 
Yongjun Chen,$^{3}$\thanks{E-mail: cyj@shao.ac.cn}
Leonid I. Gurvits,$^{2,4}$
Zsolt Paragi,$^{2}$
Aiyuan Yang,$^{5}$
Xiaolong Yang$^{3}$
\and
and Zhiqiang Shen$^{3}$
\\
$^{1}$Department of Space, Earth and Environment, Chalmers University of Technology, Onsala Space Observatory, SE-439 92 Onsala, Sweden \\
$^{2}$Joint Institute for VLBI ERIC (JIVE), Oude Hoogeveensedijk 4, 7991 PD Dwingeloo, The Netherlands \\
$^{3}$Shanghai Astronomical Observatory, Key Laboratory of Radio Astronomy, Chinese Academy of Sciences, 200030 Shanghai, China \\
$^{4}$Department of Astrodynamics and Space Missions, Delft University of Technology, Kluyverweg 1, 2629 HS Delft, The Netherlands \\
$^{5}$Max-Planck-Insitut f\"ur Radioastronomie, Auf dem H\"ugel 69, 53121 Bonn, Germany \\
}

\date{Accepted XXX. Received YYY; in original form ZZZ}

\pubyear{2021}

\begin{document}
\label{firstpage}
\pagerange{\pageref{firstpage}--\pageref{lastpage}}
\maketitle
\begin{abstract}
In the time domain, the radio sky in particular along the Galactic plane direction may vary significantly because of various energetic activities associated with stars, stellar and supermassive black holes. Using multi-epoch Very Large Array surveys of the Galactic plane at 5.0~GHz, Becker et al. (2010) presented a catalogue of 39 variable radio sources in the flux density range 1--70~mJy. To probe their radio structures and spectra, we observed 17 sources with the very-long-baseline interferometric (VLBI) imaging technique and collected additional multi-frequency data from the literature. We detected all of the sources at 5 GHz with the Westerbork Synthesis Radio Telescope, but only \target{} with the European VLBI Network (EVN). Together with its decadal variability and multi-frequency radio spectrum, we interpret it as an extragalactic peaked-spectrum source with a size of $\la$10~pc. The remaining sources were resolved out by the long baselines of the EVN because of either strong scatter broadening at the Galactic latitude $<$1~deg or intrinsically very extended structures on centi-arcsec scales. According to their spectral and structural properties, we find that the sample has a diverse nature. We notice two young H~\textsc{II} regions and spot a radio star and a candidate planetary nebula. The rest of the sources are very likely associated with radio active galactic nuclei (AGN). Two of them also displays arcsec-scale faint jet activity. The sample study indicates that AGN are commonplace even among variable radio sources in the Galactic plane.  
\end{abstract}




\section{Introduction}
\label{sec:intro}

Variable radio sources including slow transients represent rarely-seen high-energy astrophysical processes. Their counterparts might be associated with active galactic nuclei (AGN), coronally active stars, accretion activity of the compact objects in binary systems \citep[e.g.][]{Pietka2015, Mooley2016, Zhao2020}. Because they often have significant non-thermal radio emission, they are also important targets for very long baseline interferometry (VLBI) imaging observations to answer some fundamental questions. With the VLBI observations at centimetre wavelengths, we can probe their structures with a resolution of $\sim$1~mas and an astrometry precision of $\sim$10 $\mu$as \citep[e.g. ][]{Reid2014, Yang2016, Rioja2020}. For nearby Galactic variable sources, the VLBI technique allows us to measure their distances and proper motions precisely \citep[e.g.][]{Atri2020, Ding2020}, and reveals their structure changes on relatively short timescales \citep[e.g.][]{Chomiuk2014}. 

Currently, multi-epoch radio sky surveys are becoming reachable and bringing us a more complete view of the dynamic radio sky. With the 36 dual-polarization beams, it took just 300 hr for the full 36-antenna Australian Square Kilometre Array Pathfinder (ASKAP) telescope to finish the sky survey below Declination $+$40~deg \citep{McConnell2020}. The Rapid ASKAP Continuum Survey (RACS) has recently found 23 radio stars with significant circular polarization in the southern sky \citep{Pritchard2021}. The ASKAP Variables and Slow Transients (VAST) pilot survey has revealed a highly-polarized transient point source \citep{Wang2021} and 28 highly variable sources including several pulsars and radio stars \citep{Murphy2021}. The ongoing Karl G. Jansky Very Large Array Sky Survey (VLASS) at 2--4~GHz plans to have three epochs separated by 32 months \citep{Lacy2020, Gordon2021}. The VLASS at the first epoch has enabled the discoveries of a transient source consistent with a merger-triggered core collapse supernova \citep{Dong2021}, a decades-long extragalactic radio transient \citep{Law2018}, and a sample of quasars that brightened by a factor of 2--25 on decadal timescales \citep{Nyland2020}. In the low-frequency regime 100--230~MHz, \citet{Ross2021} have identified several hundred sources with significant spectral variability over a year-long timescale. According to the existing sky surveys, the fraction of variable radio sources and transients with a variability of $\ga$30 per cent on timescales of $\la$10~yr and flux densities $>$0.1~mJy is about a few percent \citep[e.g.][]{Mooley2016, Murphy2021}.  

The Galactic plane is of particular interest for various sky surveys to not only probe various Galactic objects \citep[e.g.][]{Brunthaler2021}, but also reveal rarely-seen variable sources. The VLA observations with an image sensitivity of $\sim$10 $\mu$Jy\,beam$^{-1}$ at 5.5~GHz show that the fraction of variable sources can reach $\sim$74 percent at the Galactic centre \citep{Zhao2020}. Using the historical VLA surveys \citep{Becker1994, White2005, Purcell2013} of the Galactic plane (Galactic longitude: 20--45 deg, Galactic latitude: $<$1~deg) at three epochs (1990+, 2005, 2006), \citet{Becker2010} has compiled a catalogue of 39 variable radio sources. The multi-epoch VLA observations at 5~GHz show that these variable sources are not clearly resolved with a resolution of 1.5~arcsec in the pilot study \citep{Becker2010} of Co-Ordinated Radio and Infrared Survey for High-mass star formation \citep[CORNISH\footnote{\url{https://cornish.leeds.ac.uk/public/index.php}},][]{Hoare2012, Purcell2013} and more than 50 per cent of them varied by a factor of $>$2 over a timescale of about 15~yr. These variable sources have a sky density of $\sim$1.6~per deg$^2$. In the flux density range 1--100~mJy, the sky density is twice higher than observed in extragalactic radio sources \citep[][]{deVries2004}. Because of the high sky density, a significant fraction of the catalogue are expected to be Galactic sources \citep{Becker2010}.   

\citet{Becker2010} found that many variable sources ($\sim$40 per cent) have flat or inverted radio spectra based on non-simultaneous measurements \citep[Multi-Array Galactic Plane Imaging Survey, MAGPIS][]{Helfand2006} between 1.4 and 5~GHz. None of these sources has an optical counterpart \citep{Becker2010}. There are only 18 per cent of the sources coincident with mid-IR sources \citep{Price2001, Benjamin2003, Carey2009} and 5 per cent of the sources associated with X-ray sources \citep{Sugizaki2001, Hands2004}.  Several sources are likely to be associated with star-forming regions known from infrared and millimetre surveys \citep{Price2001, Benjamin2003, Aguirre2011}. For the entire catalogue, the absence of counterparts at other wavelengths might be due to intrinsically low flux densities instead of the heavily obscuration and absorption by the Galactic medium. The nature of the remaining 80 per cent variable sources was unknown. Among the known variable sources \citep[e.g.][]{Pietka2015}, coronally active radio-emitting stars \citep[e.g. RS~Canum Venaticorum stars,][]{Gudel2002, Pritchard2021}, radio pulsars including magnetars \citep[e.g.][]{Kaspi2017}, or extragalactic transients \citep[e.g. supernovae,][]{Weiler2002} have been tentatively excluded by \citet{Becker2010} in view of their high variability, long life ($\sim$15~yr) and relatively high flux density (0.5--70~mJy). The fact that these sources are variable and mainly detected in radio domain warrants their further investigation on the sub-arcsecond angular scales and the longer timescales. To date, there are also some new sky surveys available, e.g. the RACS \citep{Pritchard2021} and the VLASS \citep{Lacy2020}, for us to provide additional temporal and spectral information on the radio properties of these variable sources. These variable sources with the flat or inverted radio spectra are also candidate radio AGN \citep[e.g.][]{Orienti2020, ODea2021}. Some radio AGN have long been known to vary intrinsically by factors of a few or more due to intrinsic effects \citep[e.g.][]{Hovatta2007, Nyland2020, Wolowska2021} and propagation effects at lower observing frequencies \citep[e.g.][]{Ross2021}. 

From the catalogue presented by \citet{Becker2010}, we selected 17 sources for the European VLBI Network (EVN) observations at 5~GHz to explore their radio structures. These sources in Table~\ref{tab:sample} were selected mainly because they have nearby compact calibrators available \citep[e.g.][]{Immer2011} for us to run phase-referencing observations efficiently and reliably \citep[e.g.][]{Beasley1995, Reid2014}. The sample also included the source G22.7194$-$0.1939, which was identified by \citep{Becker2010} as an object of special interest because its position is close to the centre of a supernova remnant.

The paper is organised as follows. We describe our radio observations and data reduction in Section~\ref{sec:obs}, and present the results observed with the EVN, the Westerbork Synthesis Radio Telescope (WSRT), the VLA and the Very Long Baseline Array (VLBA) in Section~\ref{sec:results}. We discuss the nature of the only detected source \target{}, the effects of scatter broadening on the observed structures, and the classifications of these sources in Section~\ref{sec:discussion}. We give our conclusions in the last section. Through the paper, we define the spectral index $\alpha$ using the power-law spectrum $S(\nu) \propto \nu^{\alpha}$.

\begin{table}
\caption{The 17 variable sources selected for the EVN observations at 5 GHz. Each column gives (1) source name, (2) right ascension, (3) declination and (4) VLBI phase-referencing calibrator. }
\label{tab:sample}
\centering
\setlength{\tabcolsep}{5pt}
\begin{tabular}{ccccc}
\hline
Source             & Right ascension                 & Declination                    & Calibrator \\
                   & (J2000)                         & (J2000)                        &            \\
(1)                & (2)                             & (3)                            & (4)        \\
\hline
G22.7194$-$0.1939  & $18^{\rm h}33^{\rm m}21\fs033$  & $-09\degr10\arcmin06\farcs43$  &  J1825$-$0737  \\
G22.9116$-$0.2878  & $18^{\rm h}34^{\rm m}02\fs837$  & $-09\degr02\arcmin28\farcs03$  &  J1825$-$0737  \\
G22.9743$-$0.3920  & $18^{\rm h}34^{\rm m}32\fs343$  & $-09\degr02\arcmin00\farcs69$  &  J1825$-$0737  \\
G23.4186$+$0.0090  & $18^{\rm h}33^{\rm m}55\fs619$  & $-08\degr27\arcmin15\farcs92$  &  J1825$-$0737  \\
G23.5585$-$0.3241  & $18^{\rm h}35^{\rm m}23\fs014$  & $-08\degr29\arcmin01\farcs42$  &  J1825$-$0737  \\
G23.6644$-$0.0372  & $18^{\rm h}34^{\rm m}33\fs050$  & $-08\degr15\arcmin27\farcs10$  &  J1825$-$0737  \\
G24.3367$-$0.1574  & $18^{\rm h}36^{\rm m}13\fs893$  & $-07\degr42\arcmin57\farcs64$  &  J1825$-$0737  \\
G24.5343$-$0.1020  & $18^{\rm h}36^{\rm m}23\fs987$  & $-07\degr30\arcmin54\farcs26$  &  J1825$-$0737  \\
G24.5405$-$0.1377  & $18^{\rm h}36^{\rm m}32\fs343$  & $-07\degr31\arcmin33\farcs52$  &  J1825$-$0737  \\
G25.2048$+$0.1251  & $18^{\rm h}36^{\rm m}49\fs735$  & $-06\degr48\arcmin54\farcs63$  &  J1825$-$0737  \\
G25.4920$-$0.3476  & $18^{\rm h}39^{\rm m}03\fs094$  & $-06\degr46\arcmin37\farcs38$  &  J1825$-$0737  \\
G25.7156$+$0.0488  & $18^{\rm h}38^{\rm m}02\fs785$  & $-06\degr23\arcmin47\farcs29$  &  J1825$-$0737  \\ 
G26.0526$-$0.2426  & $18^{\rm h}39^{\rm m}42\fs626$  & $-06\degr13\arcmin50\farcs28$  &  J1846$-$0651  \\
G26.2818$+$0.2312  & $18^{\rm h}38^{\rm m}26\fs372$  & $-05\degr48\arcmin35\farcs17$  &  J1846$-$0651  \\ 
G37.7347$-$0.1126  & $19^{\rm h}00^{\rm m}36\fs987$  & $+04\degr13\arcmin18\farcs60$  &  J1907$+$0127  \\
G37.7596$-$0.1001  & $19^{\rm h}00^{\rm m}37\fs037$  & $+04\degr14\arcmin59\farcs08$  &  J1907$+$0127  \\
G39.1105$-$0.0160  & $19^{\rm h}02^{\rm m}47\fs984$  & $+05\degr29\arcmin21\farcs52$  &  J1907$+$0127  \\
\hline
\end{tabular}
\end{table}

\section{Observations and data reduction}
\label{sec:obs}

\subsection{The EVN experiments}

With the EVN at 5.0~GHz, we observed the sample on 2010 December 15 and 16. We also carried out the follow-up observations of \target{} on 2012 February 8 and September 17. The experiment configurations are summarised in Table~\ref{tab:exp}. The participating stations were Effelsberg (\texttt{EF}), phased-up array (\texttt{WB}) of the WSRT,  Jodrell Bank Mark~\textsc{ii} (\texttt{JB2}), Hartebeesthoek (\texttt{HH}), Onsala (\texttt{ON}), Noto (\texttt{NT}),  Medicina (\texttt{MC}), Toru\'n (\texttt{TR}), Yebes (\texttt{YS}) and Cambridge (\texttt{CM}). The three experiments used a data rate of 1024~Mbps (16~MHz filters, 2~bit quantization, 16~sub-bands in dual polarization) in the $e$-VLBI mode \citep{Szomoru2008}. The data were transferred from the European stations to Joint Institute for VLBI ERIC (European Research Infrastructure Consortium) via broad-band internet connections and then correlated in the real-time mode by the EVN Mark\,\textsc{iv} data processor \citep{Schilizzi2001} in 2010 and the EVN software correlator \citep[\textsc{sfxc},][]{Keimpema2015} in 2012. The correlations were done with ordinary parameters for continuum experiments (1 or 2~s integration time, 16 or 32 points per subband). 

\begin{table*}
\caption{The EVN observation configurations. The two-letter code for each VLBI station is explained in Section~\ref{sec:obs}. }
\label{tab:exp}
\begin{tabular}{lccccccc}
\hline
Project  & Freq. 
                 & Bandwidth 
                         & Starting \texttt{UT} & Duration  &  Participating VLBI stations                 & Phase         &  Sun dist.   \\
 code    & (GHz) & (MHz) &                      &  (h)      &                                              & calibration   &  ($\degr$)      \\
\hline     
 EY014   & 5.004  & 128   & 2010 Dec 15, 13h    &  11      & \texttt{EF, JB2, MC, ON, TR, WB, YS, HH, CM}  & Acceptable   & 21.1  \\
EY017A   & 4.990  & 128   & 2012 Feb 08, 06h    &  6       & \texttt{EF, JB2, MC, ON, TR, WB, YS}          & Successful    & 41.8  \\
EY017B   & 4.990  & 128   & 2012 Sep 17, 15h    &  7       & \texttt{EF, JB2, MC, ON, TR, WB, YS, NT}      & Successful    & 103.1 \\
\hline
\end{tabular}
\end{table*}

The VLBI observations were performed in the phase-referencing mode \citep[e.g.][]{Beasley1995}. The source positions \citep{Becker2010} and the phase-referencing calibrators \citep{Charlot2020} are listed in Table~\ref{tab:sample}. The angular separations between each pair of sources are $\la$4$\degr$. The cycle time was $\sim$4 minutes ($\sim$60~s for calibrator, $\sim$120~s for target, $\sim$60~s for scan gaps. Each source was observed for 6--8 cycles. This gave us an actual image sensitivity of about 0.2~mJy\,beam$^{-1}$. We also observed the very bright ($>$10~Jy) calibrator 3C~454.3 as a fringe finder for a few short scans during each experiment.    

During the observations of \target{} on 2012 February 8 and September 17, we also observed a faint calibrator J1831$-$0756 as the secondary phase-referencing calibrator. The faint calibrator is just 0$\fdg$9 apart from the target and allow us to run another round of phase-referencing calibration to further remove residual phase errors.  

We calibrated the visibility data with the National Radio Astronomy Observatory (NRAO) software package: Astronomical Image Processing System \citep[\textsc{aips},][]{Greisen2003}. Firstly, we inspected the visibility data and performed standard editing and flagging. Secondly, we did an a-priori amplitude calibration with the \texttt{antab} file, which saved antenna system temperatures and gain curves.  Thirdly, we corrected the ionospheric dispersive delays via the task \texttt{TECOR}, the phase errors due to the antenna parallactic angle variations via the task \texttt{CLCOR}, and instrumental phase and delay offsets across subbands via the task \texttt{FRING}. Fourthly,  we ran the fringe-fitting on the data combining all the subbands and the Stokes $RR$ and $LL$ correlations, and applied the solutions to all the sources. Finally, the instrumental bandpass shapes were corrected via the task \texttt{BPASS}. All \textsc{aips} tasks were run via the \textsc{parseltongue} interface \citep{Kettenis2006}. 

\begin{figure}
\centering
\includegraphics[width=0.48\textwidth]{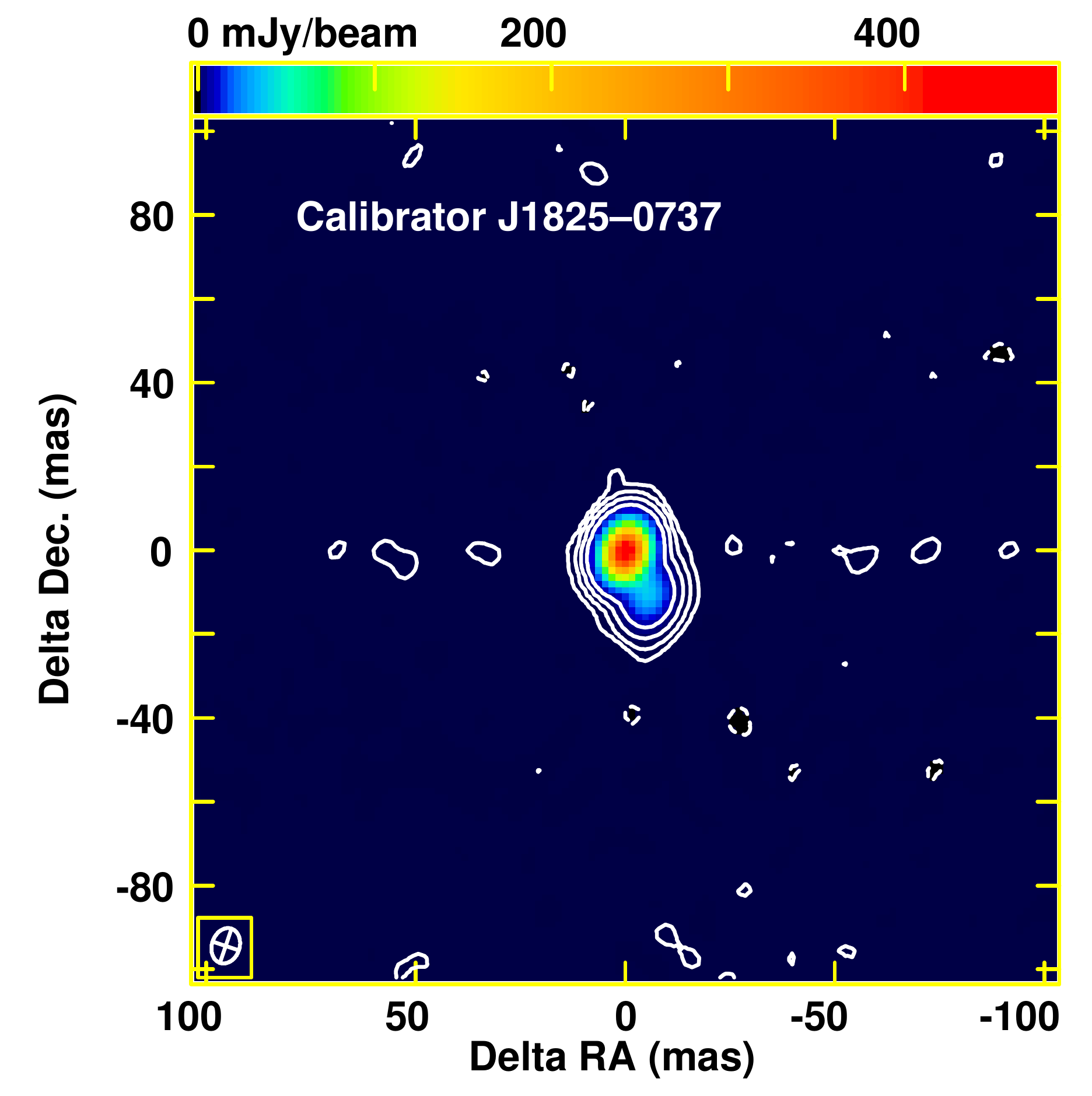}  \\
\caption{
The jet structure of the bright phase-referencing calibrator J1825$-$0737 at 5~GHz on 2012 September 17. The map is made with natural weighting. The sizes of the beam FWHM are $8.6 \times 6.6$~mas at PA~=~$-$18$\fdg$5. The map has a peak brightness of 484~mJy~beam$^{-1}$ and a noise level of $\sigma = 0.028$~mJy~beam$^{-1}$. The contours represent the level of 3$\sigma$~$\times$~($-$1, 1, 4, 16, 64). The image origin is at RA = 18$^{\rm h}$25$^{\rm m}$37$\fs$60955, Dec. = $-$07$\degr$37$\arcmin$30$\farcs$0130 (J2000). }
\label{fig:j1825}
\end{figure}

\begin{figure}
\centering
\includegraphics[width=0.46\textwidth]{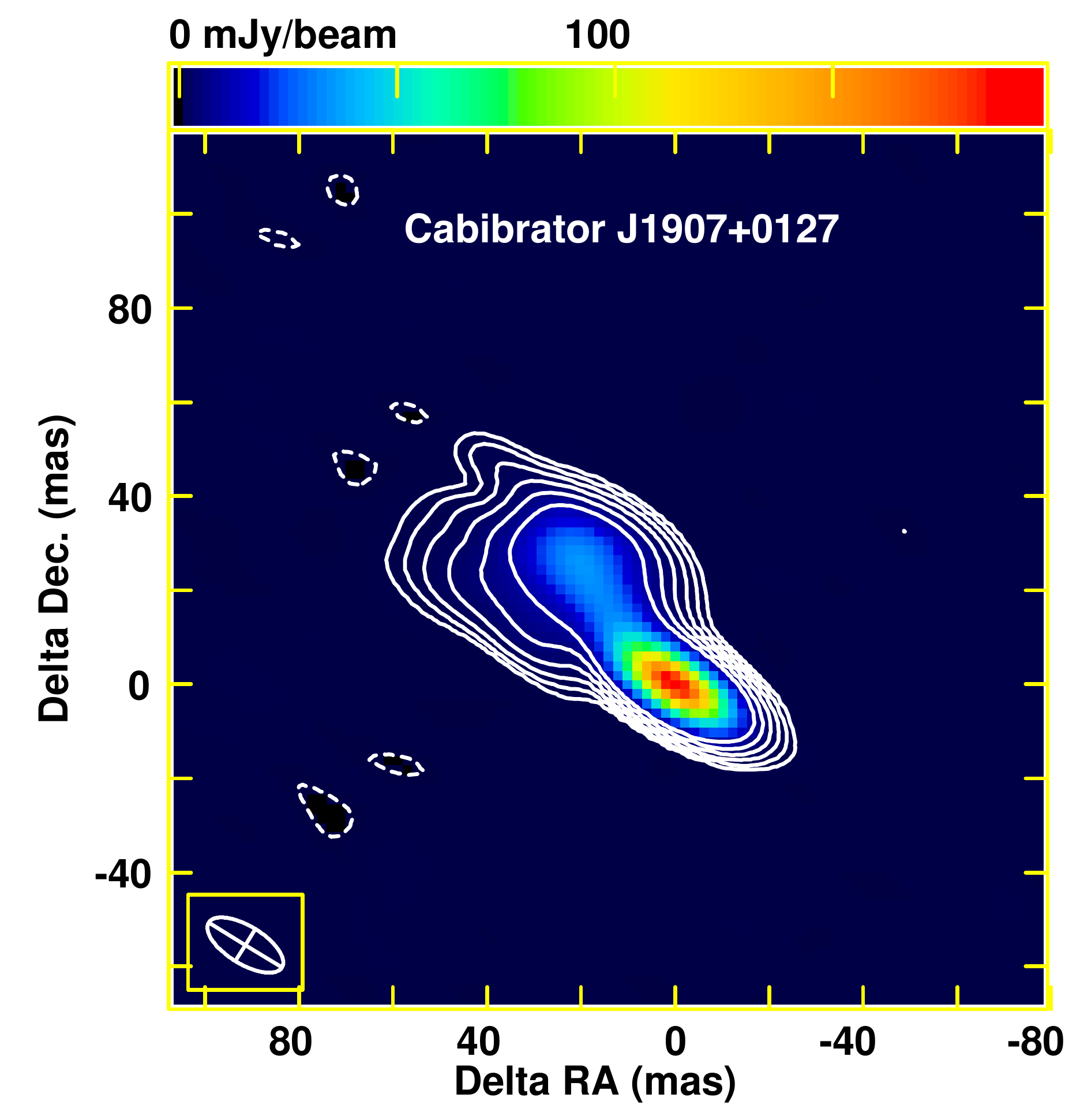}  \\
\caption{
The core-jet structure of the phase-referencing calibrator J1907$+$0127 at 5~GHz. The map is made with natural weighting. The beam sizes are $18.3 \times 8.05$~mas at PA~=~$-$58$\fdg$5. The map has a peak brightness of 197~mJy~beam$^{-1}$ and a noise level of $\sigma = 0.05$~mJy~beam$^{-1}$. The contours are at the levels 3$\sigma$~$\times$~($-$1, 1, 2, 4, 8, 16, 32, 64). The image origin is at RA = 19$^{\rm h}$07$^{\rm m}$11$\fs$99625, Dec. = $+$01$\degr$27$\arcmin$08$\farcs$9624 (J2000). }
\label{fig:j1907}
\end{figure}

\begin{figure}
\centering
\includegraphics[width=0.46\textwidth]{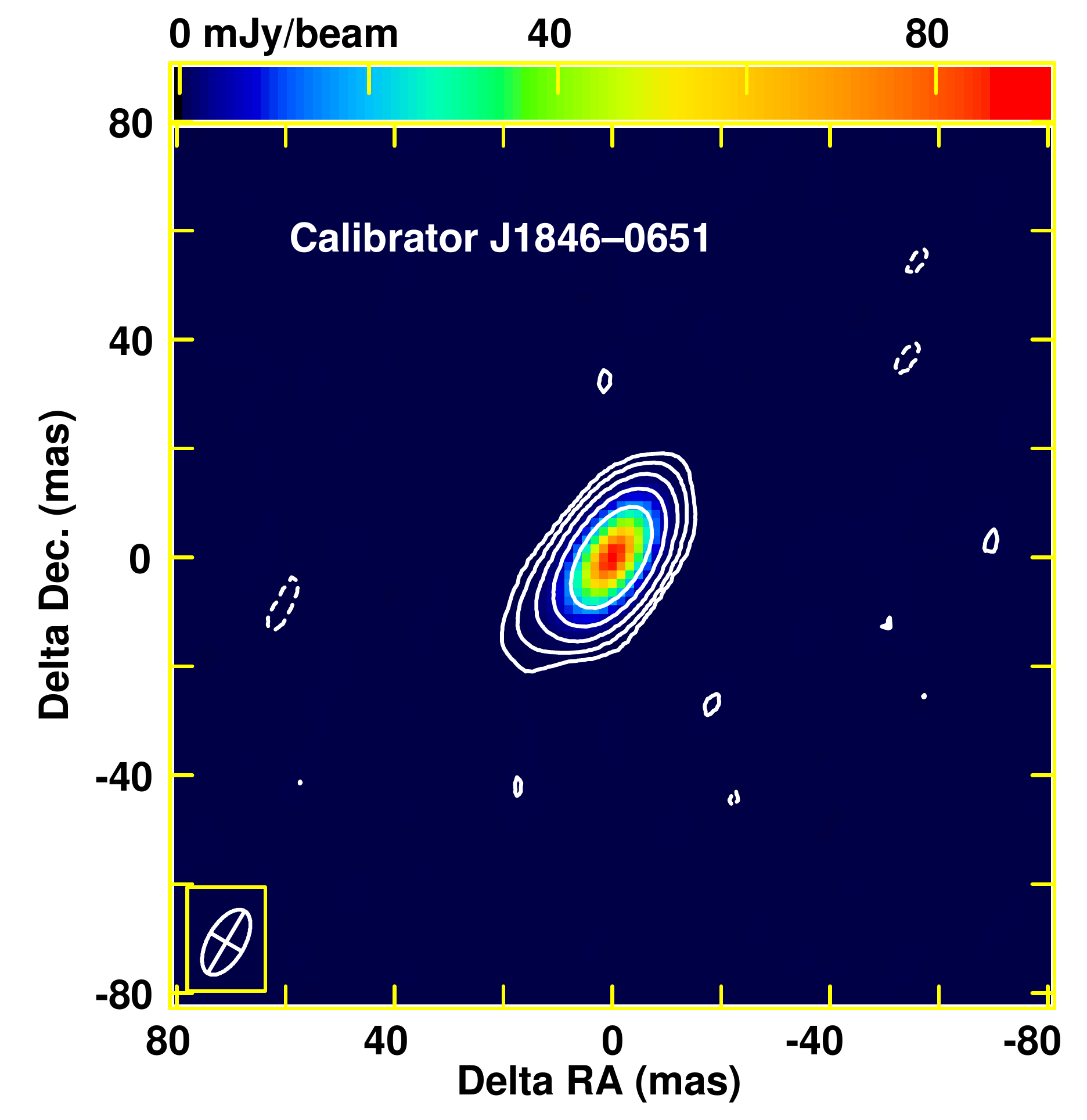}  \\
\caption{
The slightly resolved jet structure of the phase-referencing calibrator J1846$-$0651. The map is made at 5~GHz with natural weighting. The sizes of the beam FWHM are $13.3 \times 6.72$~mas at PA~=~$-$30$\fdg$4. The map has a peak brightness of 91.6~mJy~beam$^{-1}$ and a noise level of $\sigma = 0.024$~mJy~beam$^{-1}$. The contours are at the levels 3$\sigma$~$\times$~($-$1, 1, 4, 16, 64, 256). The image origin is at RA = 18$^{\rm h}$46$^{\rm m}$06$\fs$30027, Dec. = $-$06$\degr$51$\arcmin$27$\farcs$7460 (J2000). }
\label{fig:j1846}
\end{figure}

Using the calibrated visibility data, we imaged all the sources in \textsc{difmap} \citep{Shepherd1994}. The phase-referencing calibrators were imaged through iterations of fitting the un-grided visibility data to some delta functions, i.e., point source models, and running self-calibration. During each iteration, new point-source models were manually added at the peak position in the residual map. Because this method ensures the applicability of the least-squares fitting for the whole iteration, it can allow us to gain a nearly random noise distribution in the residual map even in case of a poor ($u$, $v$) coverage and a heterogeneous VLBI network \citep[e.g.][]{Yang2020imbh,Yang2021pds}. The bright calibrator J1825$-$0737 had an integrated flux density of $0.56 \pm 0.03$~Jy at 5.0~GHz in all three epochs. Fig.~\ref{fig:j1825} shows its one-sided core-jet structure. The calibrator J1907$+$0127 displays a core-jet structure with a total flux density of $0.32 \pm 0.02$~Jy at 5~GHz in Fig.~\ref{fig:j1907}. The calibrator J1846$-$0651 shows a slightly resolved core-jet morphology with a total flux density of $0.101 \pm 0.005$~Jy at 5~GHz in Fig.~\ref{fig:j1846}. The three calibrator images are consistent with the exiting VLBI survey images \citep{Charlot2020, Petrov2021}. 

\begin{figure}
\centering
\includegraphics[width=0.48\textwidth]{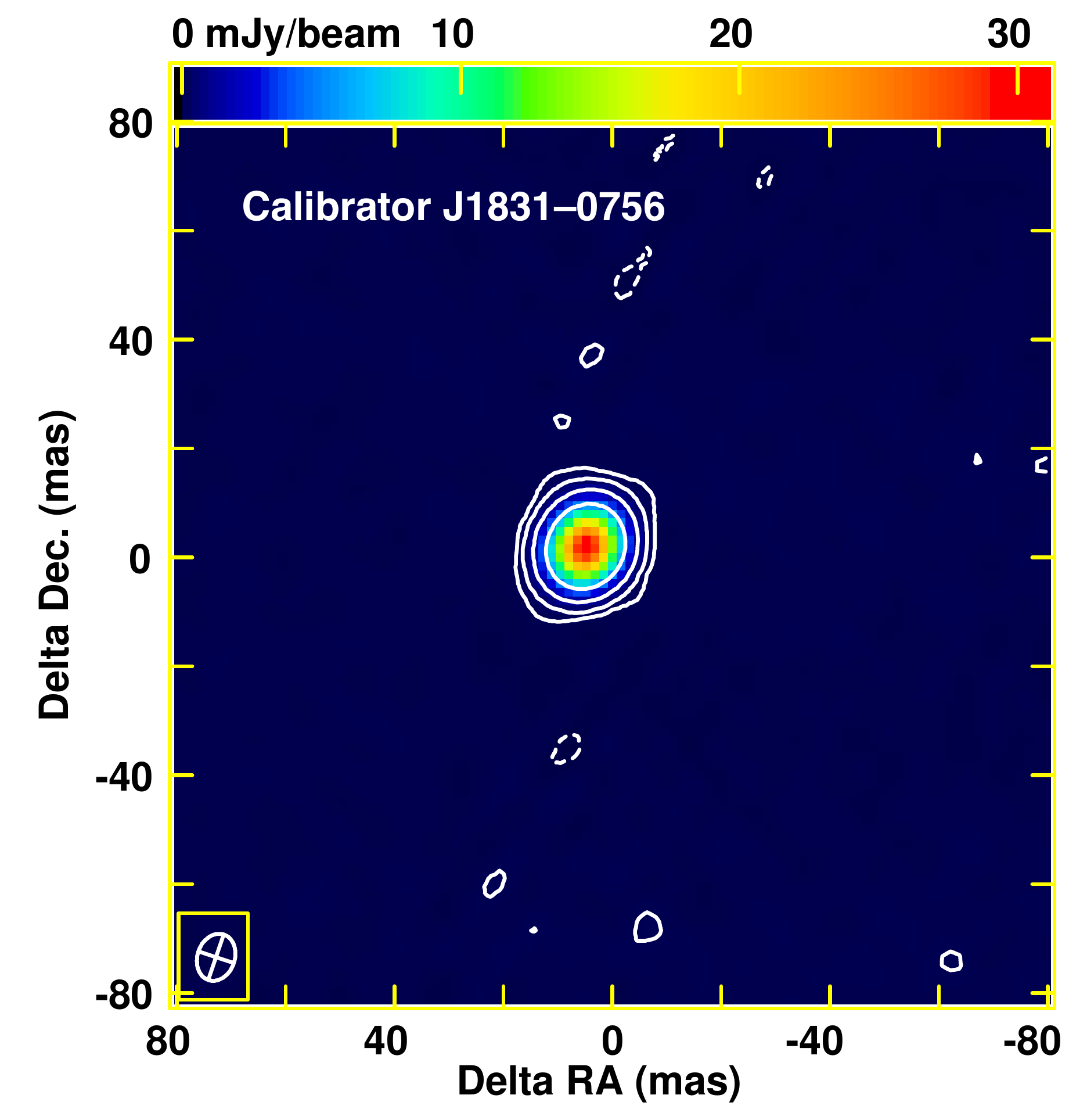}  \\
\caption{
The isotropically resolved radio morphology of the nearby faint phase-referencing calibrator J1831$-$0756 (G23.540034$+$0.871604) at 5~GHz on 2012 September 17. The map is made with natural weighting. The sizes of the beam FWHM are $8.8 \times 6.8$~mas at PA~=~$-$18$\fdg$7. The map has a peak brightness of 31~mJy~beam$^{-1}$ and a noise level of $\sigma = 0.027$~mJy~beam$^{-1}$. The contours are at the levels 3$\sigma$~$\times$~($-$1, 1, 4, 16, 64). The image origin is at RA = 18$^{\rm h}$31$^{\rm m}$03$\fs$67402, Dec. = $-$07$\degr$56$\arcmin$54$\farcs$1711 (J2000). }
\label{fig:j1831}
\end{figure}

The faint nearby calibrator J1831$-$0756 had an integrated flux density of $44 \pm 3$~mJy. Fig.~\ref{fig:j1831} displays its radio morphology, which is apparently resolved (size: $\sim$5.0~mas) and shows a very isotropic structure most likely because of the angular broadening by scattering in the Galactic interstellar medium. The correlation amplitude on the longest baseline \texttt{ON--NT} ($\sim$35~Mega-wavelengths) was $\sim$10~mJy. 

With the input calibrator images, we re-ran the fringe-fitting and the amplitude and phase self-calibration in \textsc{aips}. All these amplitude and phase solutions were also transferred to the target data using a linear interpolation. The target sources were imaged without self-calibration. The observations on 2010 December 15 had a relatively small angular distance (about 21~deg) to the Sun, and suffered an unstable ionosphere.  Thus, the fringe phase varied rapidly. This caused a relatively poor phase interpolation on the long baselines. However, on the short baselines, phase variation between neighbour scans was usually $<$180~deg and thus still acceptable for us to do an unambiguous phase interpolation.    

\subsection{The WSRT observations}
We also processed the data obtained simultaneously with our VLBI run by the WSRT on 2010 December 15. The observations were performed with a linear array of ten 25-m telescopes deployed with a maximum baseline length of 2.7~km. The northern east-west array had an observing elevation of $\le$30~deg and gave us a typical image resolution of $\sim$80~arcsec by $\sim$4~arcsec at 5~GHz with natural weighting for these low-Declination sources. During the observations, we observed 3C~48 as the primary flux density calibrator for two scans. After the labels of the linear polarisation products were changed, the Stokes $XX$ and $YY$ correlation data were calibrated in the standard way recommended by the online \textsc{aips} cookbook\footnote{\url{http://www.aips.nrao.edu/cook.html}} in appendix A. During the data reduction, we used the model of \citep{Perley2017} to calculate the flux density (5.4~Jy at 5~GHz) of 3C~48, and scaled the flux densities of the phase-referencing calibrators. In the ($u,v$) plane, we ran point-source model fitting to measure flux densities directly in \textsc{difmap}. The output flux densities and the total uncertainties (including 5 per cent as systematic errors) are reported in Table~\ref{tab:size}. 

\subsection{Additional VLA and VLBA data}

By chance, four sources were also observed with the VLA (project code 17A--070) C configuration at X band on 2017 June 2. These sources are listed in Table~\ref{tab:vla017a-070}. Because they have rising spectra between 1.4 and 5~GHz \citep{Yang2019HII}, they were selected for the VLA observations to search for Galactic thermal sources. The VLA observations covered the band 8--12 GHz continuously to measure their spectra. The calibrator 3C~286 was observed as the primary flux density calibrator \citep{Perley2017}. The sources J1851$-$0035 and J1832$-$1035 were observed as the phase-referencing calibrators. Each target source had an on-source time of $\sim$1 min. The data reduction was performed using the Common Astronomy Software Applications package \citep[\textsc{casa}][]{McMullin2007}. With the VLA pipeline\footnote{\url{https://science.nrao.edu/facilities/vla/data-processing/pipeline}}, the data were calibrated and processed. 

We notice that there were also VLBA observations of \target{} (project code BH208M) at 8.4~GHz on 2015 August 10 (private communication with Pikky Atri, James Miller-Jones and Arash Bahramian). The observations had a recording data rate of 2048 Mbps (32 MHz filters, 2 bit quantization, 16 subbands in dual polarization). \target was observed for one hour in the phase-referencing mode with a cycle time of $\sim$3~min ($\sim$30 s for J1825$-$0737, $\sim$120 s for \target{}, $\sim$30 s for two scan gaps). Following the VLBA continuum data calibration method recommended by the \textsc{aips} cookbook, we reduced the data. 

\begin{table*}
\caption{Constraints on the angular sizes of the 17 Galactic plane sources. Each column gives (1) source number, (2) name, (3) total flux density measured by the simultaneous WSRT observations at 5~GHz on 2010 December 15, i.e. epoch 2011.0 (4) constraint on the correlation amplitude on the sensitive short baseline \texttt{EF-WB} (about three Mega-wavelengths), (5) constraint on the source angular size based on the correlation amplitudes, (6) angular size of the major axis reported by the 5-GHz VLA survey CORNISH \citep{Hoare2012, Purcell2013} without de-convolution, (7) fractional variability between the 5-GHz flux densities observed by the WSRT at the epoch 2011.0 and the VLA at the epoch 2006.6 \citep{Becker2010}, (8) comments. }

\label{tab:size}
\centering
\begin{tabular}{cccccccc}
\hline
No. & Source              &$S_\textsc{2011}$ 
                                         & $S_\textsc{3mw}$  & $\theta_\textsc{fwhm}$                  
                                                                & $\theta_\textsc{vla5}$  
                                                                                       & $V_{\rm f}$ at 5~GHz
                                                                                                        & Comments  \\
    &                     & (WSRT)      & (\texttt{EF--WB}) & (EVN)   & (CORNISH)      & (WSRT vs VLA)  &           \\                                                                  
    &                     &   (mJy)       & (mJy)          & (mas)    & (arcsec)       &                &        \\
(1) & (2)                 &   (3)         & (4)            & (5)      & (6)            & (7)            & (8)    \\
\hline
1   & G22.7194$-$0.1939   & $4.8\pm0.4$   & $\leq$1.8      & $\geq$35 & $1.50\pm0.14$  & $+0.15\pm0.13$ &  Candidate radio AGN \\
2   & G22.9116$-$0.2878   & $15.6\pm0.9$  & $\leq$0.8      & $\geq$62 & $2.18\pm0.10$  & $-0.05\pm0.08$ &  Candidate radio AGN, 4.5 arcsec extension  \\
3   & G22.9743$-$0.3920   & $8.2\pm0.7$   & $\leq$1.5      & $\geq$47 & $1.50\pm0.08$  & $+0.13\pm0.10$ &  Candidate radio AGN \\
4   & G23.4186$+$0.0090   & $4.2\pm0.4$   & $\leq$0.8      & $\geq$47 & $1.64\pm0.12$  & $-0.28\pm0.12$ &  Candidate radio AGN \\
5   & G23.5585$-$0.3241   & $5.8\pm0.5$   & $\leq$2.1      & $\geq$36 & $1.90\pm0.15$  & $+0.06\pm0.11$ &  Candidate radio AGN \\
6   & G23.6644$-$0.0372   & $23.8\pm1.3$  &      12.8      &       28 & $1.50\pm0.04$  & $-0.09\pm0.09$ &  VLBI detection, radio AGN \\
7   & G24.3367$-$0.1574   & $6.9\pm0.5$   & $\leq$1.0      & $\geq$50 & $1.66\pm0.11$  & $+0.01\pm0.11$ &  Candidate radio AGN \\
8   & G24.5343$-$0.1020   & $8.5\pm0.7$   & $\leq$0.9      & $\geq$54 & $1.50\pm0.09$  & $+0.61\pm0.11$ &  Radio star \\
9   & G24.5405$-$0.1377   & $7.3\pm0.7$   & $\leq$0.9      & $\geq$52 & $1.50\pm0.11$  & $+0.47\pm0.13$ &  Candidate radio AGN \\
10  & G25.2048$+$0.1251   & $6.5\pm0.5$   & $\leq$0.9      & $\geq$51 & $1.92\pm0.19$  & $+0.43\pm0.12$ &  Planetary nebula or radio AGN   \\ 
11  & G25.4920$-$0.3476   & $4.9\pm0.5$   & $\leq$0.8      & $\geq$49 & $1.50\pm0.50$  & $+0.84\pm0.17$ &  Candidate radio AGN \\ 
12  & G25.7156$+$0.0488   & $16.5\pm0.9$  & $\leq$0.8      & $\geq$63 & $2.35\pm0.03$  & $+0.27\pm0.08$ &  H~\textsc{ii} region \\ 
13  & G26.0526$-$0.2426   & $13.3\pm0.8$  & $\leq$2.5      & $\geq$47 & $1.50\pm0.06$  & $+0.05\pm0.08$ &  Candidate radio AGN \\
14  & G26.2818$+$0.2312   & $19.6\pm1.1$  & $\leq$0.7      & $\geq$66 & $1.50\pm0.04$  & $+0.25\pm0.08$ &  Candidate radio AGN \\ 
15  & G37.7347$-$0.1126   & $12.2\pm0.8$  & $\leq$0.8      & $\geq$60 & $1.78\pm0.05$  & $+0.00\pm0.09$ &  H~\textsc{ii} region \\
16  & G37.7596$-$0.1001   & $13.0\pm0.9$  & $\leq$0.7      & $\geq$62 & $1.50\pm0.05$  & $+0.08\pm0.09$ &  Candidate radio AGN, 12.5 arcsec extension  \\
17  & G39.1105$-$0.0160   & $4.1\pm0.5$   & $\leq$1.1      & $\geq$41 & $2.24\pm0.32$  & $-0.02\pm0.15$ &  Candidate radio AGN \\
\hline
\end{tabular}
\end{table*}

\begin{table*}
\caption{Total flux densities of the 17 variable sources. Columns give  (1) source number, (2) name (the superscript $^*$: Galactic star or H~\textsc{ii} region), (3) flux density from the ASKAP survey RACS at 887.5~MHz between 2019 and 2020 \citep{McConnell2020}, (4--6) flux densities from the VLA survey THOR at 1.06, 1.44 and 1.95 GHz in 2013 \citep{Bihr2016},  (7) flux density from the survey VLASS between 2019 Mar 10 and 2020 Aug 12 \citep{Lacy2020},  (8) simultaneous spectral index measured by the survey THOR, and (9) non-simultaneous spectral index between 0.9 and 3.0~GHz. }
\label{tab:flux}
\centering
\begin{tabular}{lcccccccc}
\hline
No. & Source                &  $S_{0.88}$  & $S_{1.06}$   & $S_{1.44}$   & $S_{1.95}$   & $S_{3.00}$   & $\alpha_{1.0}^{2.0}$  
                                                                                                                        & $\alpha_{0.9}^{3.0}$ 
                                                                                                                                         \\
    &                       & (mJy)        & (mJy)        & (mJy)        & (mJy)        & (mJy)        &                &                \\
(1) &  (2)                  & (3)          & (4)          & (5)          & (6)          & (7)          & (8)            & (9)            \\
\hline
1   & G22.7194$-$0.1939     & $16.2\pm1.6$ & $18.0\pm3.1$ & $10.7\pm1.6$ &  $7.2\pm1.3$ &  $6.4\pm0.7$ & $-1.68\pm0.36$ & $-0.77\pm0.12$ \\
2   & G22.9116$-$0.2878     & $72.1\pm5.5$ & $55.6\pm2.8$ & $47.4\pm2.0$ & $32.2\pm1.7$ & $28.8\pm1.3$ & $-0.96\pm0.09$ & $-0.75\pm0.07$ \\
3   & G22.9743$-$0.3920     & $13.1\pm1.4$ & $10.1\pm5.1$ &  $9.5\pm2.8$ &  $5.7\pm2.0$ & $10.3\pm0.9$ & $-1.31\pm2.71$ & $-0.20\pm0.11$ \\
4   & G23.4186$+$0.0090     & $\leq1.3$    & \nodata      & \nodata      & \nodata      &  $4.9\pm0.6$ & \nodata        & $\geq1.1$      \\
5   & G23.5585$-$0.3241     & $15.6\pm1.6$ & $14.5\pm2.4$ & $11.2\pm1.5$ & \nodata      &  $8.1\pm0.8$ & $-0.59\pm0.45$ & $-0.54\pm0.12$ \\
6   & G23.6644$-$0.0372     & $\leq1.3$    & \nodata      & \nodata      & \nodata      & $13.5\pm0.9$ & \nodata        & $\geq1.9$      \\
7   & G24.3367$-$0.1574     & $23.9\pm2.2$ & $25.5\pm0.3$ & $18.9\pm2.1$ & \nodata      & $12.5\pm0.9$ & $-0.54\pm0.26$ & $-0.53\pm0.10$ \\
8   & G24.5343$-$0.1020$^*$ & $\leq1.3$    & \nodata      & \nodata      & \nodata      &  $4.8\pm0.7$ & \nodata        & $\geq1.10$     \\
9   & G24.5405$-$0.1377     & $\leq1.3$    &  $2.2\pm3.4$ &  $4.9\pm2.1$ & $3.4\pm1.6$  &  $6.8\pm0.8$ & \nodata        & $\geq1.40$     \\
10  & G25.2048$+$0.1251     & $\leq1.3$    &  $4.4\pm2.2$ &  $4.8\pm1.7$ & $4.7\pm1.2$  &  $6.9\pm0.7$ & $+0.38\pm1.06$ & $\geq1.40$     \\ 
11  & G25.4920$-$0.3476     & $13.8\pm1.4$ &  $7.3\pm2.1$ &  $8.1\pm1.6$ & $2.3\pm1.9$  &  $6.2\pm0.7$ & $+0.36\pm1.08$ & $-0.65\pm0.13$ \\
12  & G25.7156$+$0.0488$^*$ & $\leq1.3$    & $12.2\pm4.2$ & $15.3\pm2.4$ & \nodata      & $21.4\pm1.4$ & $+0.58\pm0.54$ & $\geq2.3$      \\ 
13  & G26.0526$-$0.2426     & $24.7\pm2.2$ & $22.1\pm1.4$ & $20.1\pm1.2$ & \nodata      & $15.5\pm1.0$ & $-0.38\pm0.15$ & $-0.38\pm0.09$ \\
14  & G26.2818$+$0.2312     & $25.8\pm2.3$ & $25.9\pm1.4$ & $25.7\pm1.0$ & \nodata      & $18.8\pm1.0$ & $+0.00\pm0.13$ & $-0.26\pm0.09$ \\ 
15  & G37.7347$-$0.1126$^*$ & $\leq1.3$    & $15.7\pm0.4$ & $13.3\pm2.1$ & $11.3\pm1.5$ & $13.5\pm1.0$ & $-0.27\pm0.38$ & $\geq1.9$      \\
16  & G37.7596$-$0.1001     & $31.6\pm2.7$ & \nodata      & \nodata      & \nodata      & $11.5\pm1.0$ & \nodata        & $-0.83\pm0.10$ \\
17  & G39.1105$-$0.0160     & $14.3\pm1.5$ & $12.4\pm1.2$ & $12.1\pm1.0$ & $12.8\pm1.5$ &  $8.6\pm0.8$ & $-0.15\pm0.19$ & $-0.42\pm0.11$ \\
\hline
\end{tabular}
\end{table*}

\section{Results}
\label{sec:results}

\subsection{WSRT and EVN imaging results
}

\begin{figure}
\includegraphics[width=\columnwidth]{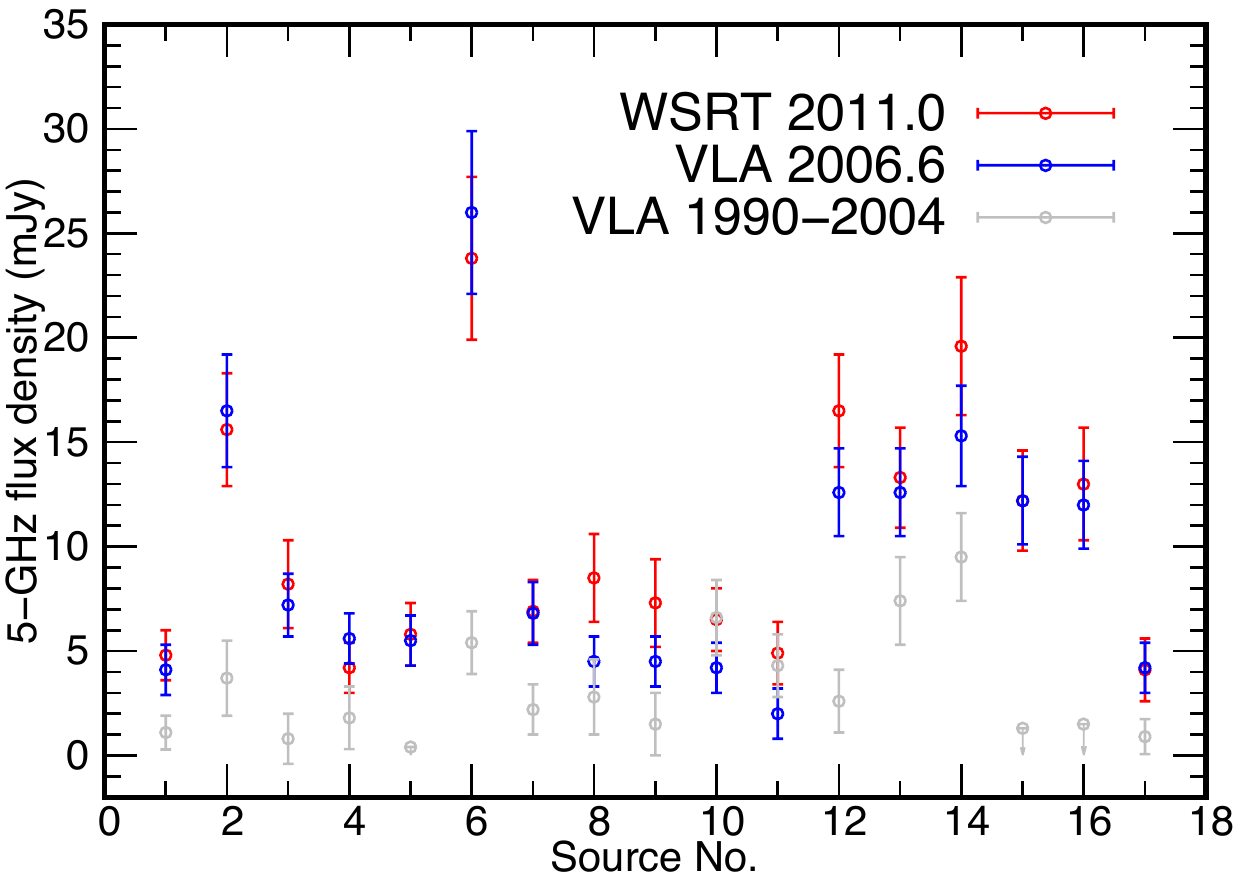} \\
\caption{
The flux densities observed at 5 GHz by the VLA at the epochs 1990$+$ and 2006.6 \citep{Becker2010} and the WSRT at the epoch 2011.0. The error bars represent 3$\sigma$. The WSRT data are listed in Table~\ref{tab:size}. We notice that the non-detections of the sources No. 5, 15 and 16 at the early epoch were because of poor deconvolution (c.f. Section~\ref{subsec:discussion_alpha}).}
\label{fig:var}
\end{figure}

Among the 17 variable radio sources, only \target{} was clearly detected in the EVN observations on 2010 December 15. Using the data on the shortest and most sensitive baseline \texttt{EF--WB} (distance 266~km), we made a dirty map and took the peak brightness as an upper limit on the average correlation amplitude for each source in Table~\ref{tab:size}. All the 17 sources were detected by the simultaneous low-resolution WSRT observations. Fig.~\ref{fig:var} displays the 5-GHz flux densities measured by the WSRT at the epoch 2011.0 and the VLA at the epochs 2006.6 and 1990$+$ \citep{Becker2010}. Over the first two epochs 1990$+$ and 2006.6, 15 of 17 sources were greatly brightened \citep{Becker2010}. To search for significant variability between the later two epochs, we calculated the fractional variability $V_{f}$, which is defined as: 
\begin{equation}
V_{\rm f} = 2 \left( \frac{S_{2011} - S_{2006}}{S_{2011} + S_{2006}} \right ).
\label{eq:Vf}
\end{equation}
The results are also listed in Table~\ref{tab:size}. Over 4.4 yr, one source (No. 8: G24.5343$-$0.1020) had a fractional variability of $>$5$\sigma$, and six sources (No. 4, 9--12, 14) had a fractional variability in the range 3$\sigma$--5$\sigma$.  

For sources with a brightness distribution following a circular Gaussian function \citep[e.g.][]{Pearson1995}, we can describe its correlation amplitude $S(r)$ as:  
\begin{equation}
S(r) = S_{0} \exp \left( \frac{-(\pi r \theta_\textsc{fwhm})^2}{4\ln2} \right),
\label{eq:theta}
\end{equation}
where $\theta_\textsc{fwhm}$ is the source FWHM in radians, $r$ is the ($u$, $v$) radius in wavelength, and $S_0$ is the correlation amplitude at $r=0$, i.e. total flux density. Using the model~\ref{eq:theta}, the measured WSRT flux density and the upper limit of the correlation amplitude on the baseline \texttt{EF--WB} at about three mega-wavelengths (M$\lambda$), we derived a lower limit on the angular size for each source. These constraints are reported in Table~\ref{tab:size}. The minimum angular size among all sources in the observed sample is 28~mas. Moreover, we searched for their angular sizes in the CORNISH catalogue \citep{Hoare2012, Purcell2013}.  The angular size of the major axis is reported in Table~\ref{tab:size}. Compared to the restoring beam size (1.5~arcsec), six sources are slightly resolved.   

These sources have been also observed at 2--4 GHz in the ongoing survey VLASS \citep{Lacy2020}. The survey has a resolution of 2.5~arcsec and a sensitivity of 0.12~mJy\,beam$^{-1}$ per epoch. Using the image cutout web service\footnote{\url{http://cutouts.cirada.ca/}} provided by the Canadian Initiative for Radio Astronomy Data Analysis (CIRADA), we downloaded their total intensity images that had the more accurate flux density calibration at the later epochs 1.2 and 2.1. During the multi-epoch VLA observations between 2019 Mar 10 and 2020 Aug 12, all the 17 sources were clearly detected. We used the \textsc{aips} task \texttt{JMFIT} to determine their integrated flux densities. The flux densities at 3.00~GHz are presented in Table~\ref{tab:flux}. We also used 5 per cent of the total flux density as the systematic errors. 

In the continuum survey RACS at 887.5~MHz \citep{McConnell2020}, ten sources from our sample were clearly detected. We also list their flux densities in Table~\ref{tab:flux}. If the source is not detected, we only give a 5$\sigma$ (the survey median value $\sigma = 0.25$~mJy\,beam$^{-1}$) upper limit on its total flux density. Using the surveys RACS and VLASS with a time interval of about one year,  we also derived a non-simultaneous spectral index between 887.5 MHz and 3.0~GHz for each source in Table~\ref{tab:flux}. Because a systematic error of $\sim$5 per cent is included in the two surveys, the spectral index estimate may be affected significantly by a fractional variability of $>$15 per cent. 

These Galactic plane sources are also observed by The HI/OH/Recombination line survey of the inner Milky Way \citep[THOR,][]{Bihr2016, Wang2018, Wang2020} in 2013. The observations with the VLA at 1--2~GHz used eight 128-MHz subbands with the centre frequencies from 1.052 to 1.948 MHz, and gained image resolutions 10--40 arcsec. The continuum source catalogue provides simultaneously flux density measurements and spectral indices. In the catalogue, we searched for our target sources with a radius of 5~arcsec and found 13 sources. We also checked their online images to further verify these detections. Their flux densities at 1.05, 1.44, and 1.95~GHz and spectral indices are reported in Table~\ref{tab:flux}. 

\subsection{The multi-epoch EVN imaging results of \target{}}

\begin{figure*}
\includegraphics[width=0.40\textwidth]{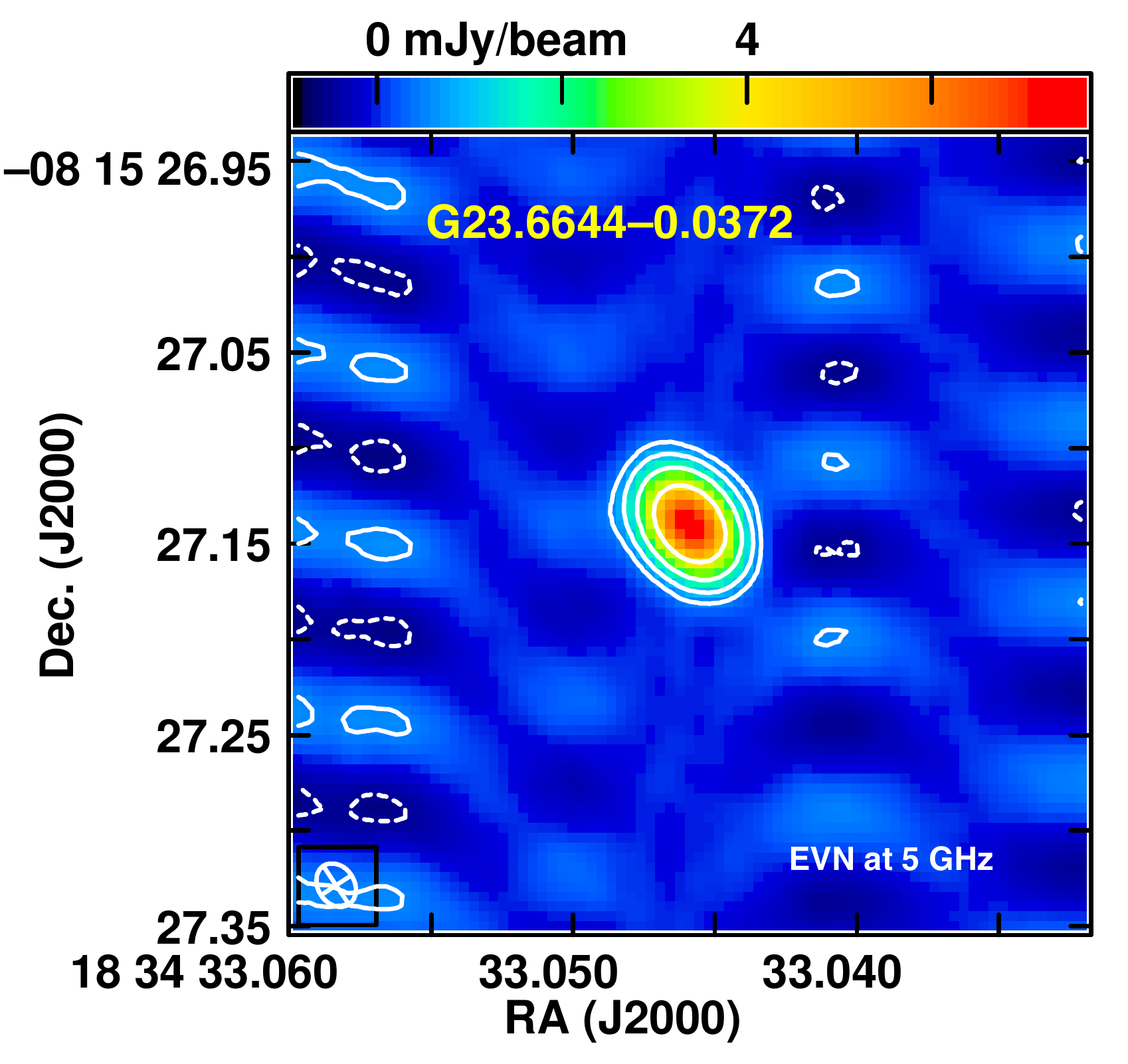}  
\includegraphics[width=0.59\textwidth]{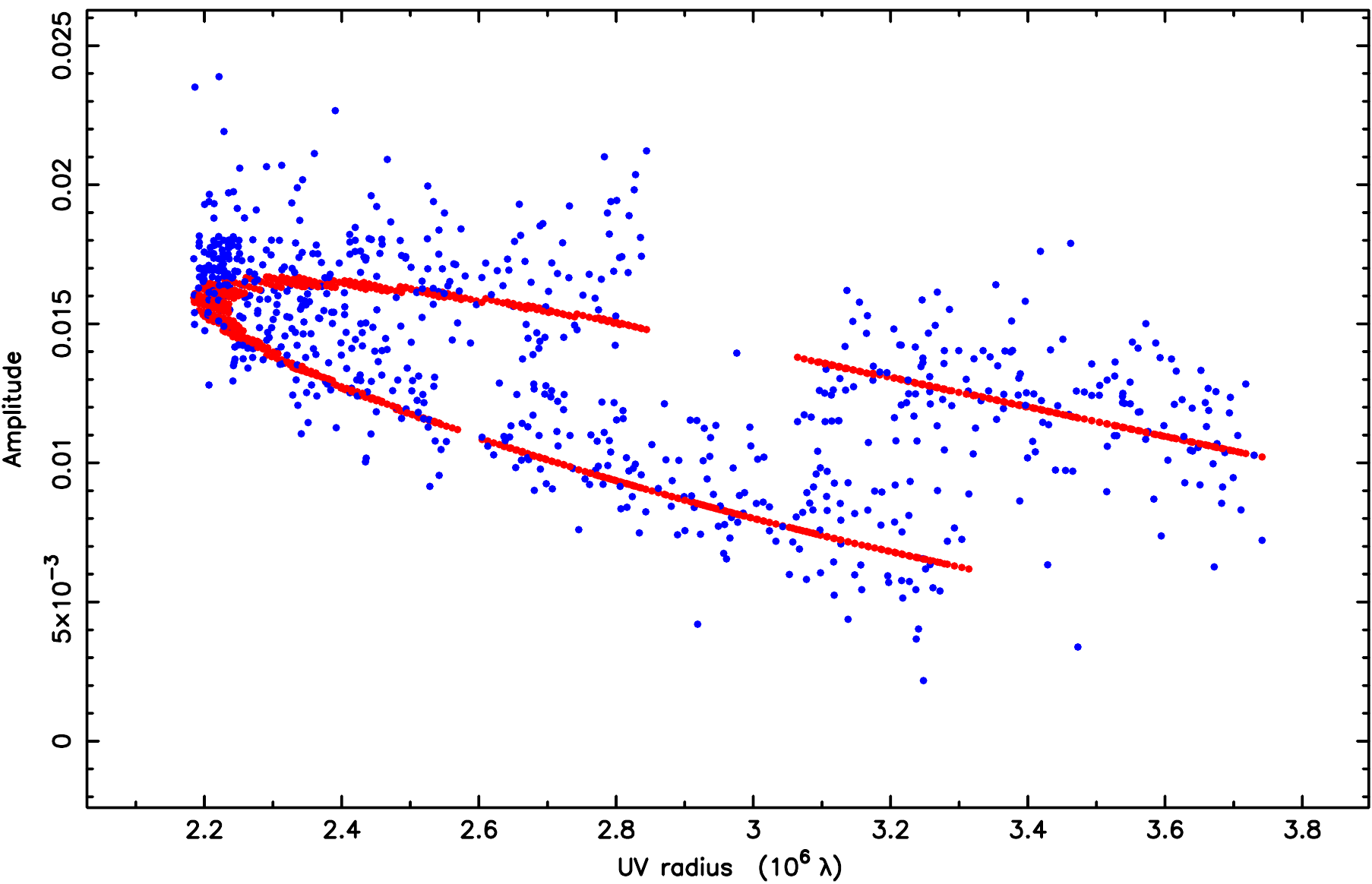}  \\
\caption{
The Stokes $I$ map and the best-fitting model of the variable radio source \target{}. 
Left: The intensity map at 5 GHz on 2012 September 17. The image has a noise level of $\sigma = 0.2$~mJy~beam$^{-1}$ and a peak brightness of 7.6~mJy~beam$^{-1}$. The sizes of the beam are FWHM~=~$24 \times 19$~mas at PA~=~33$\fdg$4. The contours represent the levels 2.5$\sigma$~$\times$~($-$1, 1, 2, 4, 8). Right: The correlation amplitude in Jy versus the ($u$, $v$) radius in Mega-wavelengths for the source \target{} on the shortest baseline of \texttt{EF--WB}. The red dots represent the best-fitting elliptical Gaussian model used in the image. The blue dots show the visibility data.}
\label{fig:g23}
\end{figure*}

\begin{figure}
\includegraphics[width=\columnwidth]{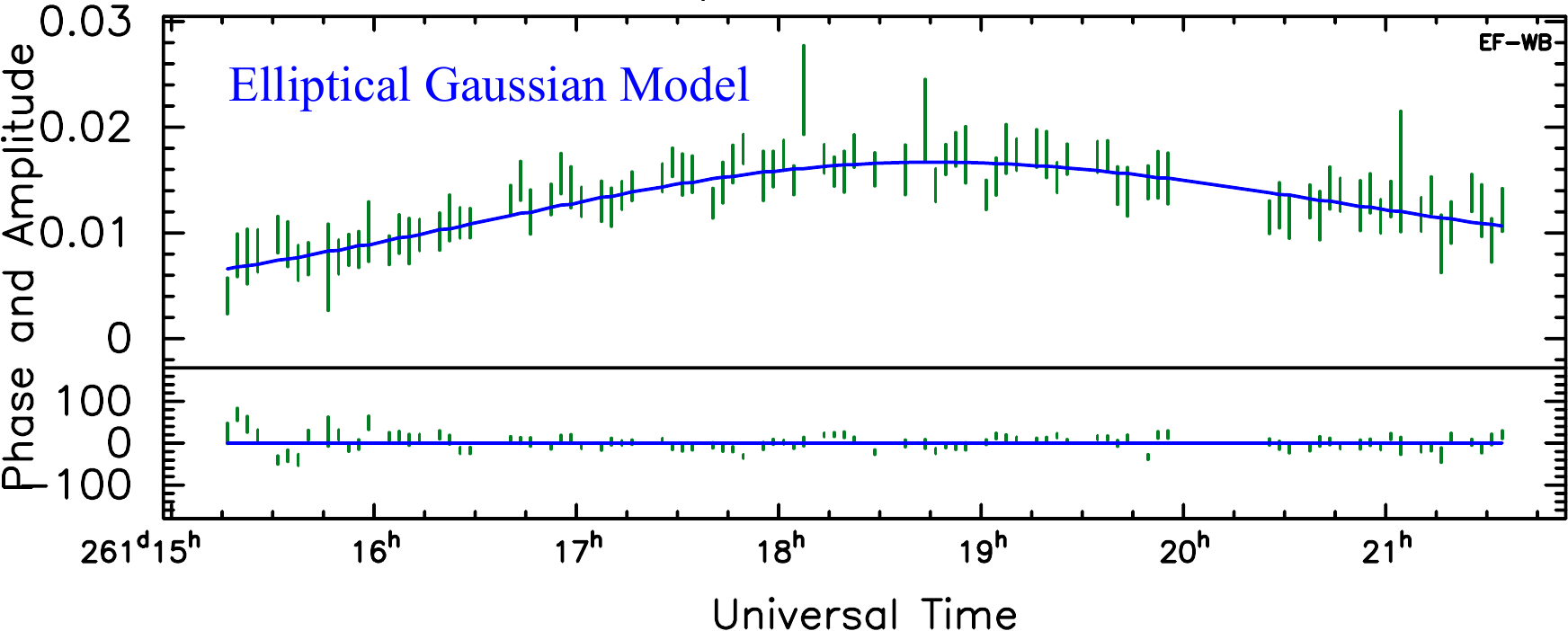}  \\
\includegraphics[width=\columnwidth]{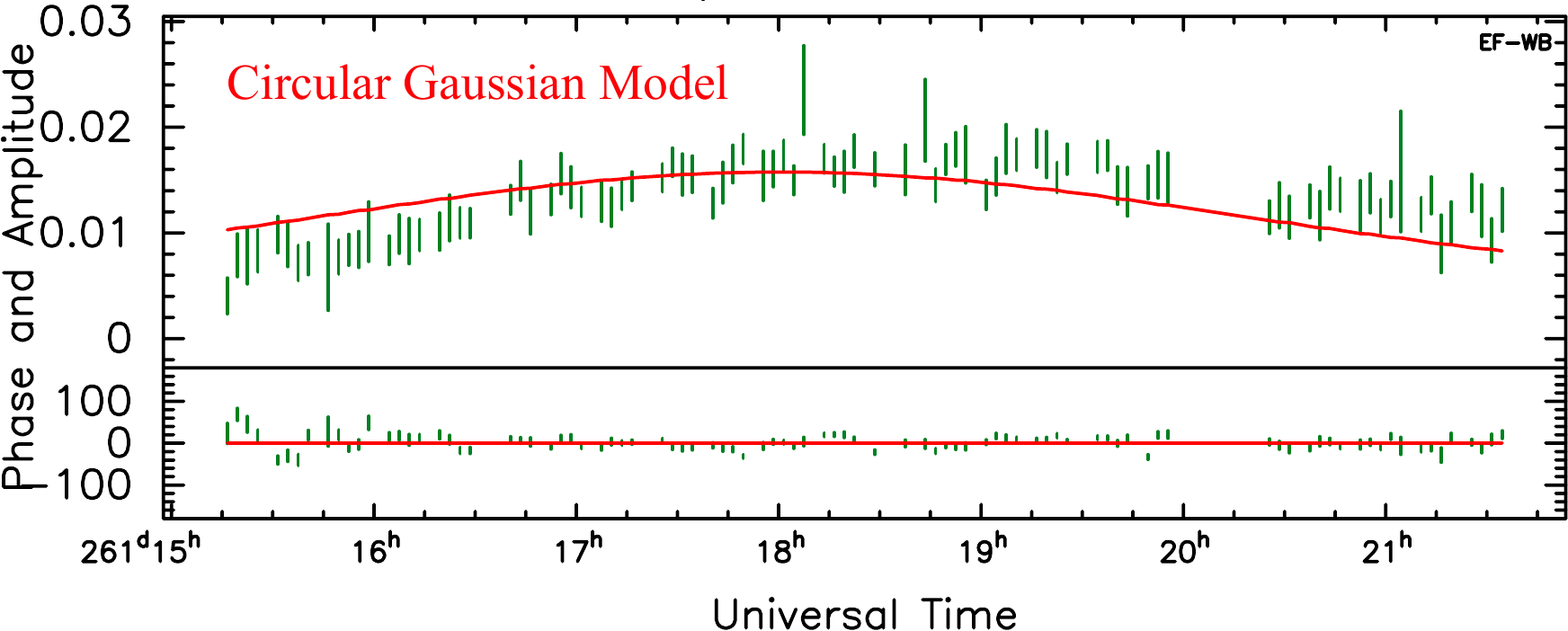} \\
\caption{
The best-fitting circular and elliptical Gaussian models and the visibility data on the short baseline \texttt{EF--WB} in the first subband. }
\label{fig:g23_model}
\end{figure}

The left panel in Fig.~\ref{fig:g23} shows the imaging results for \target{}. The Stokes $I$ image was made using the data on the short baselines $\leq$10 M$\lambda$ and natural weighting. The source does not look clearly resolved in the image plane. However, the correlation amplitude varies significantly from $\sim$15 to $\sim$5~mJy on the shortest baseline \texttt{EF--WB}, indicating that the source is significantly resolved \citep[e.g.][]{Pearson1995}. The correlation amplitude variation is also displayed in the right panel of Fig.~\ref{fig:g23}. On the other baselines, fringes are not clearly seen. Without the baseline \texttt{EF--WB}, \target{} shows a peak brightness of $0.18$~mJy~beam$^{-1}$, which is slightly below the general 5$\sigma$ detection limit in the dirty map with natural weighting. This indicates that the source has a very low surface brightness and is over-resolved \citep[e.g.][]{Pearson1995}.  We tried to fit the short-baseline visibility data to both circular and elliptical Gaussian models in \textsc{difmap}. The visibility data and the models on the baseline \texttt{EF--WB} in the first subband are shown in Fig.~\ref{fig:g23_model}. Compared to the circular Gaussian model, the elliptical Gaussian model gives a 2 per cent smaller reduced $\chi^2$ and fit the long baseline data (c.f. Fig.~\ref{fig:g23}) of\texttt{EF--WB} more reasonably. In view of the more accurate fitting, we only report the results of the elliptical Gaussian model in Table~\ref{tab:fit}. The uncertainties associated with the best-fitting parameters are the formal errors at the reduced $\chi^{2}_{\rm red} = 1$. The systematic uncertainty for the integrated flux density is $\sim$1.3~mJy ($\sim$5 per cent). We excluded the long baseline data at the radius $>10$~M$\lambda$ in the model fitting.    

\begin{figure}
\includegraphics[width=\columnwidth]{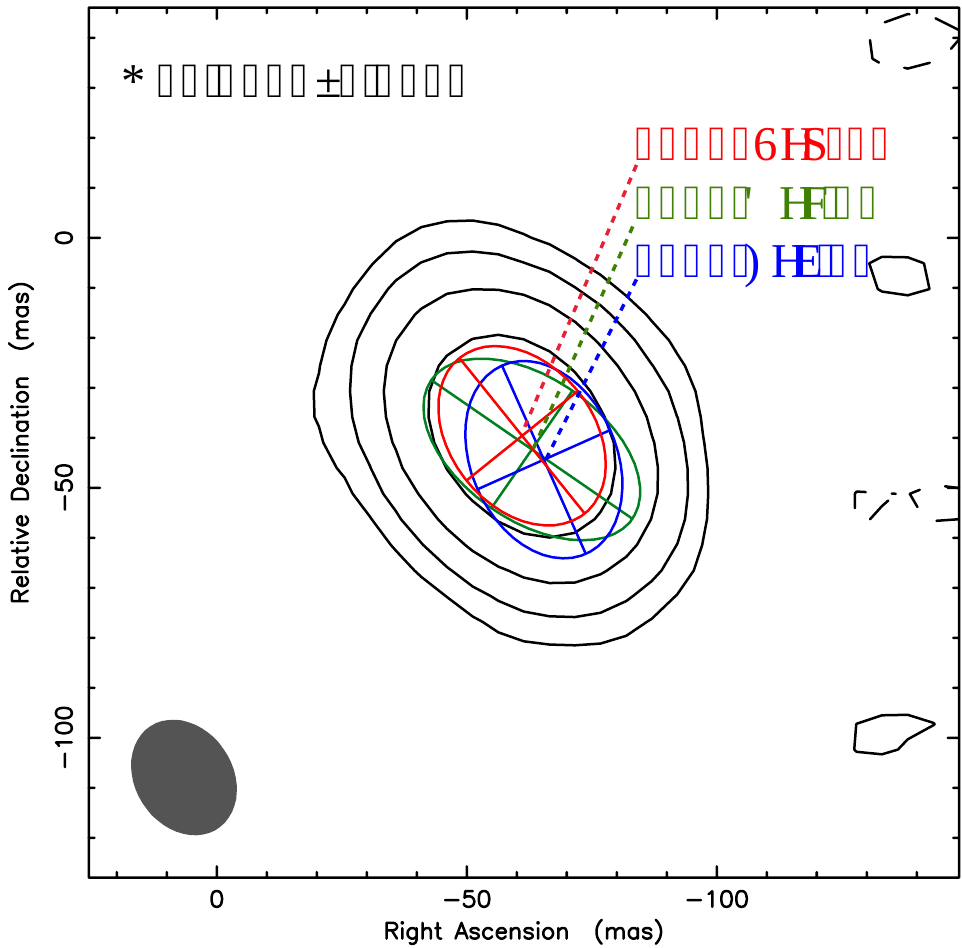} 
\caption{The multi-epoch positions and sizes of the variable radio source \target{}. The contour image in Fig.~\ref{fig:g23} is also over-plotted here. The data in details are reported in Table~\ref{tab:fit}. }
\label{fig:g23_3ep}
\end{figure}

Over the three epochs of EVN observations covering about two years, \target{} had no detectable proper motion. Fig.~\ref{fig:g23_3ep} presents the measured positions and sizes. The source centroid has a large scatter along the major extension direction. This mainly results from some systematic residual phase errors of the phase-referencing calibrations \citep{Reid2014, Rioja2020}. Our target had small angular distances from the Sun (cf. Table~\ref{tab:exp}) in the first two epochs. Because of solar activity, the interplanetary plasma and the Earth's ionosphere varied significantly from epoch to epoch. These observations also had relatively low observing elevations (15--31~deg at \texttt{EF} and \texttt{WB}).  The useful ($u$, $v$) coverage was limited to a few short baselines owning to the heavily resolved source structure. Moreover, the sensitive baseline \texttt{EF--WB} gave some strong sidelobes in the dirty map. Thus, the EVN observations failed to achieve a typical astrometric precision of $<1$~mas \citep[e.g.][]{Yang2016, Rioja2020}. The average J2000 centroid position for \target{} is RA = 18$^{\rm h}$34$^{\rm m}$33$\fs$0457, Dec. = $-$08$\degr$15$\arcmin$27$\farcs$142. Assuming a stationary centroid, we estimate the standard deviations: $\sigma = 2.2$~mas in RA, $\sigma = 2.4$~mas in Dec.  The source structure shows a consistent extension along the direction $\theta_{\rm pa}= 39.4 \pm 2.7$~deg. The average ellipsoid axes are $\theta_{\rm maj} = 43.4 \pm 0.7 $~mas and $\theta_{\rm min} = 28.7 \pm 1.3$~mas.   

The brightness temperature $T_{\rm b}$ is estimated \citep[e.g.][]{Condon1982} as
\begin{equation}
T_\mathrm{b} = 1.22\times10^{9}\frac{S_\mathrm{int}}{\nu_{\rm obs}^2\theta_{\rm maj}\theta_{\rm min}}(1+z),
\label{eq1}
\end{equation}
where $S_\mathrm{int}$ is the integrated flux density in mJy, $\nu_\mathrm{obs}$ is the observing frequency in GHz, $z$ is the redshift and unkown, $\theta_{\rm maj}$ and $\theta_{\rm min}$ are the FWHM in mas. The average brightness temperature over the three epochs will be $(1.1\pm0.1)\times10^6(1+z)$~K.  

\subsection{Imaging results of the additional VLA and VLBA data}

The broad-band VLA observations gave us a sensitivity of $\sim$0.1~mJy\,beam$^{-1}$ and a resolution up to $\sim$3~arcsec by $\sim$2~arcsec. All the four sources are clearly detected. The simultaneous multi-frequency flux density measurements are listed in Table~\ref{tab:vla017a-070}. The flux densities were measured from the sub-band data of 1~GHz bandwidth. Since some channels with poor data were flagged out, each sub-band had a valid bandwidth of about 900~MHz. For each source, we also tried to estimate the angular size by model-fitting a circular Gaussian model with the \textsc{casa} task \textsc{uvmodelfit}. However, the deconvolved angular sizes ($\leq$57 mas) are much smaller than the beam sizes and there are no significant differences between the total flux densities and the image peak values, indicating that these sources are unresolved. From the four simultaneous flux density measurements, the spectra index is also estimated and listed in Table~\ref{tab:vla017a-070}.

\begin{table*}
\caption{The VLA X-band imaging results of four variable sources with the rising spectra at $\leq$5~GHz. Each column gives (1--2) source No. and name, (3-6) flux densities at 8.448, 9.344, 10.142 and 11.038 GHz, (7) beam size at 11.038 GHz and (8) spectral index derived from the four simultaneous measurements. }
\label{tab:vla017a-070}
\centering
\begin{tabular}{cccccccc}
\hline
No. & Source             &  $S_{8.44}$        & $S_{9.34}$      & $S_{10.14}$       & $S_{11.04}$       &   Beam                      & $\alpha$\\
    &                    &  (mJy)             & (mJy)           & (mJy)             & (mJy)             &  (arcsec$\times$arcsec)     &  \\
(1) & (2)                & (3)                &  (4)            &  (5)              & (6)               & (7)                         &   (8)  \\
\hline
4   & G23.4186$+$0.0090  &  $26.6\pm1.3$     &   $26.3\pm1.3$   &   $26.1\pm1.3$    &   $25.6\pm1.3$    &  $3.09\times2.07$           & $-0.13\pm0.03$ \\
6   & G23.6644$-$0.0372  &  $30.4\pm1.5$     &   $31.9\pm1.6$   &   $32.7\pm1.7$    &   $33.3\pm1.7$    &  $3.06\times2.06$           & $+0.34\pm0.04$ \\
8   & G24.5343$-$0.1020  &  $7.2\pm0.4$      &   $8.0\pm0.4$    &   $8.6\pm0.5$     &   $8.8\pm0.4$     &  $3.03\times2.07$           & $+0.73\pm0.14$ \\
10  & G25.2048$+$0.1251  &  $7.7\pm0.4$      &    $7.3\pm0.4$   &    $7.2\pm0.4$    &   $7.4\pm0.4$     &  $2.98\times2.08$           & $-0.14\pm0.14$ \\
\hline
\end{tabular}
\end{table*}

\begin{figure}
\includegraphics[width=\columnwidth]{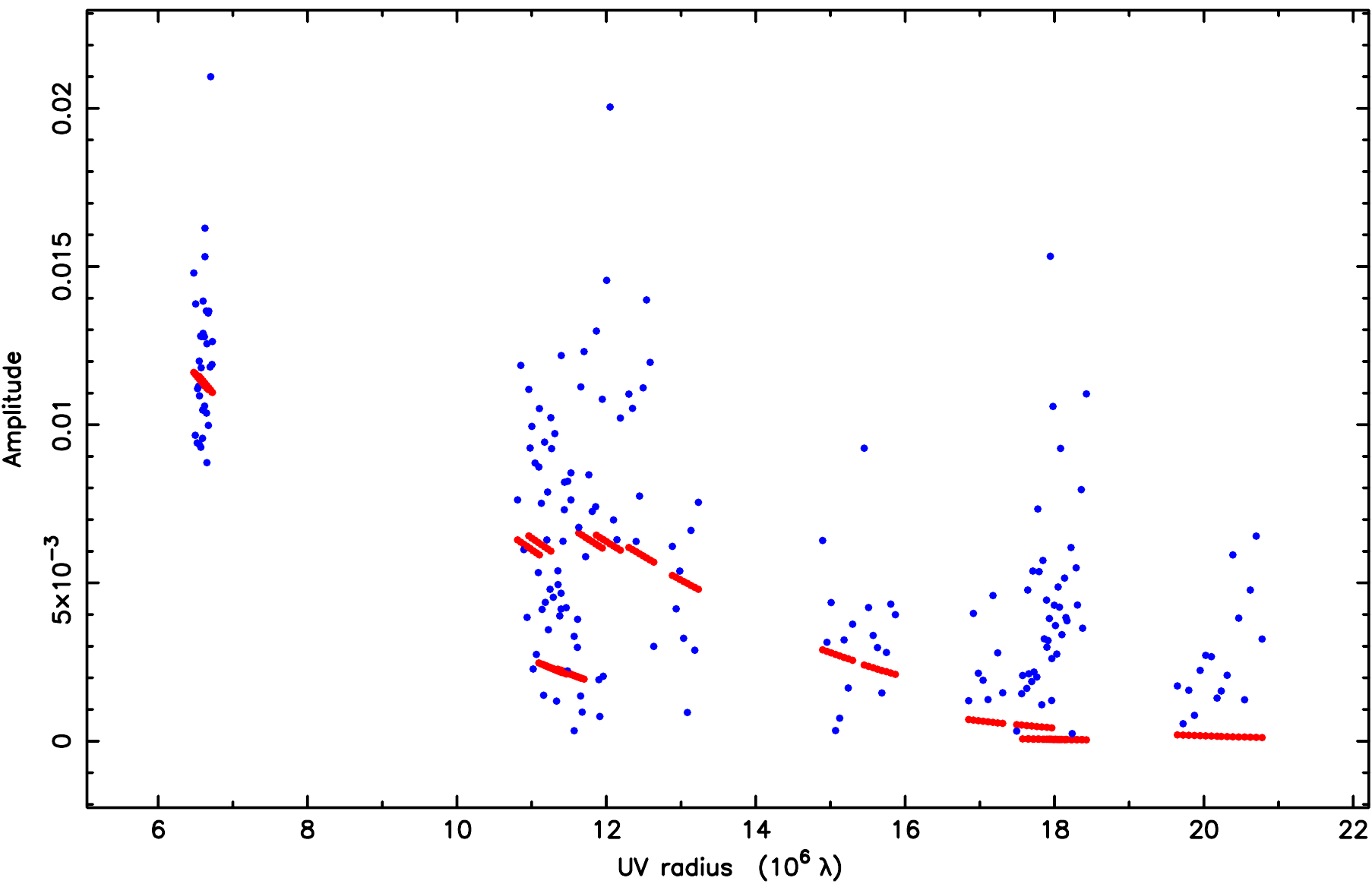} 
\caption{Plot of the 8.4-GHz correlation amplitude in Jy versus the ($u$, $v$) radius in M$\lambda$. The blue points represent the visibility data observed by the VLBA and averaged with an interval of 15~min. The red points display the best-fitting elliptical Gaussian model. The amplitude drop on the long baselines indicates that \target{} is significantly resolved. }
\label{fig:g23_x_radplot}
\end{figure}

\target{} also displays a significant resolved structure at 8.4~GHz. In Fig.~\ref{fig:g23_x_radplot}, its correlation amplitude declines significantly on the long baselines. Using an elliptical Gaussian model and the calibrated visibility data from the four short-baseline stations (Los Alamos, Pie Town, Owens Valley and Kitt Peak), we derived $S_{\rm int}=27.7\pm1.8$~mJy (including an empirical systematic error of 5 per cent), $\theta_{\rm maj} = 15.8 \pm 0.8$~mas, $\theta_{\rm min} = 10.1 \pm 1.4$~mas and $\theta_{\rm pa} = 38.0 \pm 3.4$~deg. The predicted values of the model are also plotted as the red points in Fig.~\ref{fig:g23_x_radplot}. The total flux density is consistent with that observed by the VLA (c.f. Table~\ref{tab:vla017a-070}).

\begin{table*}
\caption{Summary of the elliptical Gaussian model fitting results and map parameters of \target{}. Columns give (1) MJD, (2) total flux density, (3--4) relative offsets in Right Ascension and Declination with respect to the VLA position (RA = 18$^{\rm h}$34$^{\rm m}$33$\fs$05, Dec = $-$08$\degr$15$\arcmin$27$\farcs$1, J2000), (5--6) sizes of major and minor axes of the elliptical Gaussian model, (7) position angle, (8) map peak brightness and statistical rms, (9--11) sizes of major and minor axes of the synthesised beam, and position angle of the major axis. The errors in columns (2--7) are formal errors from the model fitting in \textsc{difmap} at the reduced $\chi_{\rm red} = 1$. }
\label{tab:fit}
\centering
\begin{tabular}{ccccccccccc}
\hline
MJD     & $S_{\rm int}$ & $\Delta$RA    &  $\Delta$Dec  & $\theta_{\rm maj}$ 
                                                                       & ${\theta_{\rm min}}$ 
                                                                                      & $\theta_{\rm pa}$ & $S_{\rm pk}$ &  $\phi_{\rm maj}$ 
                                                                                                                                 & $\phi_{\rm min}$ 
                                                                                                                                         & $\phi_{\rm pa}$ \\
        & (mJy)         & (mas)        &  (mas)        & (mas)        &  (mas)       & ($\degr$)         & (mJy\,beam$^{-1}$) 
                                                                                                                       & (mas) & (mas) &  ($\degr$) \\
 (1)    & (2)           & (3)           & (4)           & (5)          & (6)          & (7)             & (8)          & (9)   & (10)  & (11) \\ 
\hline
55546.5 & $24.5\pm2.0$  & $-63.0\pm0.5$ & $-42.3\pm0.3$ & $48.9\pm1.6$ & $28.6\pm3.0$ & $+55.4\pm5.5$   &  $5.5\pm0.9$ & 22.6  & 18.0  &  35.1 \\
55965.4 & $29.2\pm2.0$  & $-65.3\pm0.1$ & $-44.4\pm0.1$ & $41.2\pm1.2$ & $29.1\pm1.9$ & $+24.0\pm4.7$   & $10.4\pm0.4$ & 27.0  & 25.0  &  10.1 \\
56187.8 & $26.4\pm1.6$  & $-61.1\pm0.1$ & $-39.6\pm0.1$ & $39.9\pm0.6$ & $28.6\pm0.9$ & $+38.9\pm3.5$   &  $7.6\pm0.2$ & 24.0  & 19.0  &  33.4 \\ 
\hline
\end{tabular}
\end{table*}


\section{Discussion}
\label{sec:discussion}

Based on the above inputs and the available multi-wavelength data in literature, we probe the natures of the 17 sources. Firstly, the only VLBI-detected source \target{} is studied in details. Secondly, the general structural properties of the sample and the contributions of the scatter broadening are discussed. Finally, the possible nature and the spectral classification for each source are probed.

\subsection{G23.6644\texorpdfstring{$-$}{-}0.0372: a broadened image of a compact source}
\label{sec:discussion_g23}

Diffuse, turbulent and ionized gas in the Galactic interstellar medium (ISM) scatters radio waves and might cause significant temporal and angular broadening in particular at low observing frequencies and close to the Galactic plane \citep[e.g.][]{Fey1991, Cordes1991, Cordes2002}. Radio observations of intrinsically compact sources along the Galactic plane can also offer some potential for probing the medium. The spatial power spectrum of the interstellar density fluctuations is a power law,
\begin{equation}
P_\mathrm{\delta n_{e}} = C_n^2 q^{-\epsilon},
\label{eq:p}
\end{equation}
where $C_n^2$ is the spectral coefficient, $q$ is the wavenumber with $q_0$ and $q_1$ corresponding to the outer and inner scales of the turbulence, respectively. The density spectral index $\epsilon$ is empirically determined to be $\sim$11/3 \citep[e.g.][]{Cordes1991}. The scattering measure (SM) representing the line of sight integral of the power spectrum normalisation constant can be expressed as,
\begin{equation}
{\rm SM} = \int_{0}^{D} C_n^2(s) d s,
\label{eq:SM}
\end{equation}
where $D$ is the path length through the scattering medium. Because of angular broadening, the observed angular size $\theta_{\rm obs}$ follows $\theta_{\rm obs}^2 = \theta_{\rm int}^2 + \theta_{\rm scat}^2$, where $\theta_{\rm int}$ is the intrinsic source size and $\theta_{\rm scat}$ is the contribution from the scatter broadening. In case of an intrinsically compact source, i.e. $\theta_{\rm int} \ll \theta_{\rm scat}$, one can get $\theta_{\rm obs}=\theta_{\rm scat}$. Under the assumption of a uniform medium, the measurement $\theta_{\rm scat}$ can allow us to provide an estimate for SM via the relation \citep{Cordes2002, Cordes2003}:
\begin{equation}
{\rm SM} =  \left ( \frac{\theta_{\rm scat}}{\theta_0} \right )^{5/3} \nu^{11/3},
\label{eq:SMtheta}
\end{equation}
where SM is in kpc\,m$^{-20/3}$, $\nu$ is in GHz, $\theta_0 = 128$~mas for extragalactic sources and $\theta_0 = 71$~mas for Galactic sources. The equation \ref{eq:SMtheta} predicts a strong dependence of $\theta_{\rm scat} \propto \nu^{-11/5}$. 

The pulsar observations can also measure SM \citep[][]{Cordes2002, Cordes2003}:
\begin{equation}
{\rm SM} =  292 \left ( \frac{\tau_{\rm d}}{D} \right )^{5/6} \nu^{11/3},
\label{eq:SMtau}
\end{equation}
where $\tau_{\rm d}$ is the time delay because of temporal broadening in s and $D$ is the distance to the pulsar. This method only applies to pulsars because it requires very narrow pulses.

\begin{figure}
\includegraphics[width=\columnwidth]{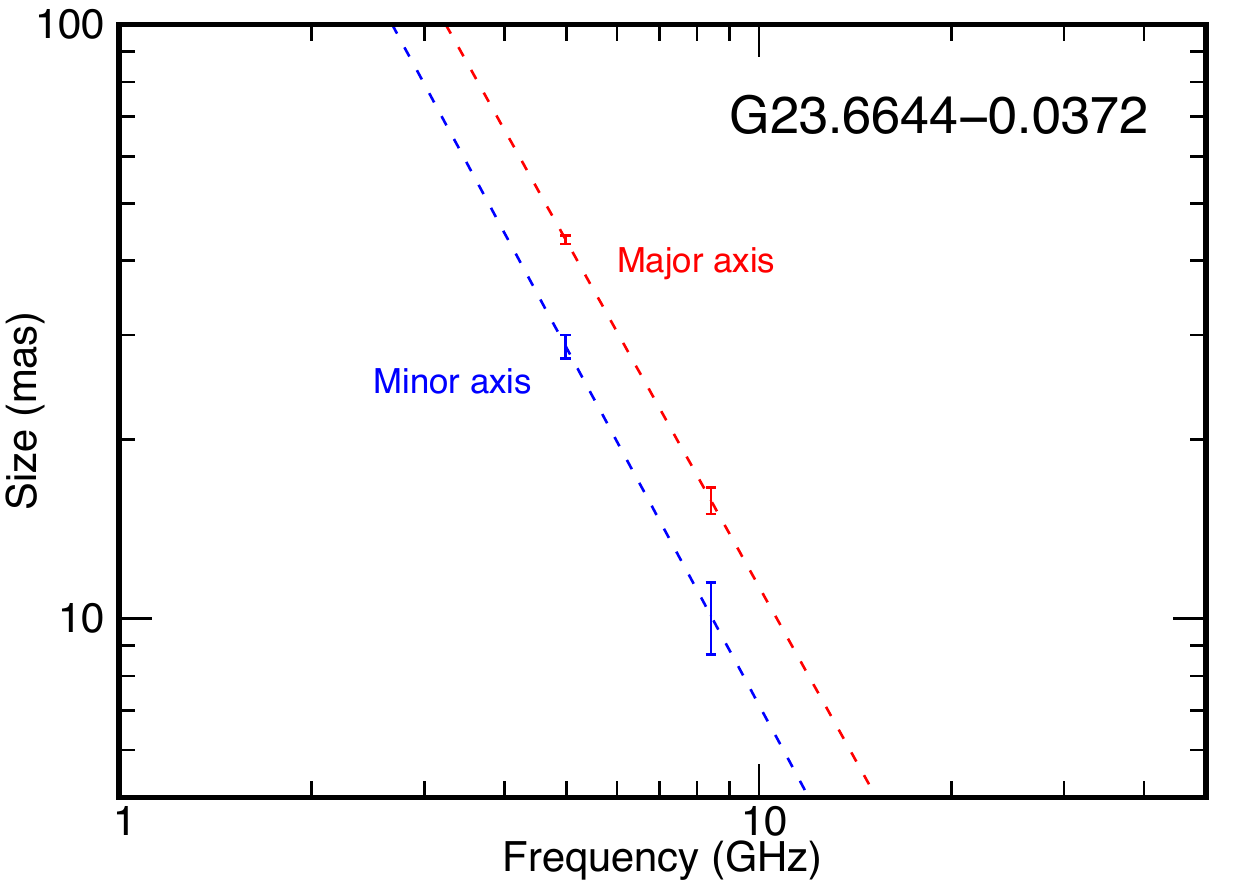} 
\caption{The best-fitting major and minor sizes of \target{} and their strong dependence on the observing frequency. The dashed lines represent the power law models: $\theta(\nu) = 970.5 \nu^{-1.93}$ (red) and $\theta(\nu) = 712.4 \nu^{-2.00}$ (blue). }
\label{fig:g23_size}
\end{figure} 

\target{} can be explained as a compact radio AGN seen in the presence of strong scattering when its radio emission propagates through the turbulent ISM of the Galactic plane. Fig.~\ref{fig:g23_size} plot the sizes of the major and minor axes at 5.0 and 8.4~GHz. It is clearly seen that the observed sizes show a strong frequency dependence. The dependence can be written as: 
\begin{eqnarray}
\theta_{\rm maj} = \theta_{\rm obs1} \left( \frac{\nu}{\nu_{\rm obs}} \right ) ^{-1.93\pm0.10}, \\
\theta_{\rm min} = \theta_{\rm obs2} \left( \frac{\nu}{\nu_{\rm obs}} \right ) ^{-2.00\pm0.28},
\label{eq:thetanu}
\end{eqnarray}
where $\theta_{\rm obs1}$ and $\theta_{\rm obs2}$ represent the observed angular sizes of the major and minor axes at the frequency $\nu_{\rm obs}$. The two power-law relations are consistent with the expectation of the scatter broadening. The derived SM is in the ranges of 30--60 kpc\,m$^{-20/3}$ for an extragalactic source and 80-160 kpc\,m$^{-20/3}$ for a galactic source. The anisotropic structure roughly along the Galactic plane direction can be produced by the source itself, the magnetic field in the scattering medium \citep[e.g.][]{Trotter1998, Kounkel2018} or both. To date, there are many similar objects reported in the literature \citep[e.g.][]{Fey1991, Pushkarev2015}. The most strongly broadened object is NGC~6334B \citep{Moran1990}, which has apparent angular sizes 0.34~arcsec by 0.27~arcsec at 4.9~GHz because of anisotropic radio-wave scattering \citep{Trotter1998}. Sagittarius A$^*$ (Sgr~A$^*$) is another case studied intensively by multi-wavelength VLBI observations \citep[e.g.][]{Shen2005, Johnson2018}. Its two-dimensional scattering structure can be described as $\theta_{\rm maj}=(1.39\pm0.02)\lambda^{2}$ mas by $\theta_{\rm min}=(0.69\pm0.06)\lambda^{2}$~mas with a position angle of $\sim$80$\degr$ \citep{Shen2005}, where $\lambda$ is in cm. The source J053424.63$-$052838.5 at the edge of the Orion nebula has a scattering disc with major by minor axes of 34~mas by 19~mas at 5~GHz \citep{Kounkel2018}. 

\begin{figure}
\includegraphics[width=\columnwidth]{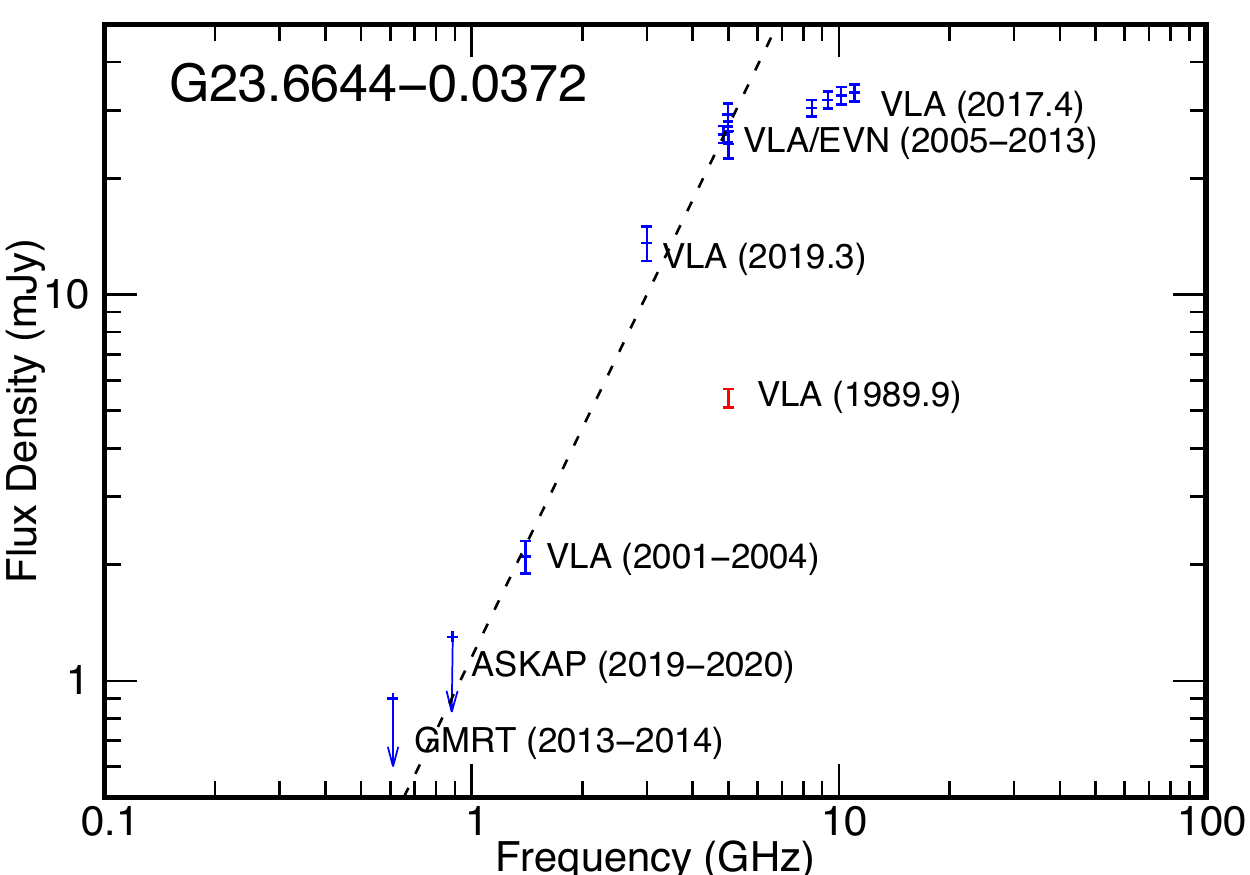} 
\caption{The rising radio spectrum of the variable radio source \target{}. The data are from various non-simultaneous observations and listed in Table~\ref{tab:g23flux}. The dashed line represents the best-fitting power law function $S(\nu)=(1.15\pm0.21) \nu^{1.97\pm0.12}$ derived from the blue points between 1.4 and 5.0~GHz. }
\label{fig:g23_alpha}
\end{figure}

\target{} suffers strong scattering because of its low Galactic latitude \citep[e.g.]{Cordes1991, Cordes2003, Pushkarev2015}. For Galactic pulsars, there is a positive correlation between SM and dispersion measure (DM). The DM is defined as:
\begin{equation}
{\rm DM} = \int_{0}^{D} n_e d s,
\label{eq:DM}
\end{equation}
where $n_e$ is the number density of free electrons. The correlation gives SM~$\propto$~DM$^3$ for DM $>$20~pc\,cm$^{-3}$ \citep[e.g.][]{Cordes2003}. According to the correlation relation, it requires DM $\ga$1000~pc\,cm$^{-3}$ for SM $\ga$10~kpc\,m$^{-20/3}$. In the field of our target, there is a normal radio pulsar J1834$-$0812: RA = 18$^{\rm h}$34$^{\rm m}$29$\fs$8(9), Dec. = $-$08$\degr$12$\arcmin$00(100), period 0.491~s \citep{Bates2011}. The pulsar has a very high DM, $1020 \pm 50$~cm$^{-3}$\,pc, a DM distance of $9.6 \pm 1.2$~kpc, an average flux density of $0.10 \pm 0.02$~mJy at 6.5~GHz and a spectral index of $-0.7 \pm 0.3$ between 1.4 and 6.5~GHz \citep{Bates2011}. Within a circle with a centre at \target{} and a radius of 5~deg, the pulsar has the highest DM in the ATNF pulsar catalogue\footnote{\url{https://www.atnf.csiro.au/research/pulsar/psrcat/}}. The source \target{} is within the large 3$\sigma$ error ellipse (semi-major and semi-minor axes: 400 arcsec by 300 arcsec) of PSR~J1834$-$0812. However, \target{} has a rising radio spectrum at frequencies $\leq5$~GHz(cf. Fig.~\ref{fig:g23_alpha} and discussion below). This spectrum cannot allow us to explain \target{} as an optically thin pulsar nebula associated with PSR~J1834$-$0812.

\begin{table}
\caption{Summary of the flux densities of \target{}. }
\label{tab:g23flux}
\centering
\begin{tabular}{ccc}
\hline
Freq.      & Flux density &  Array and date (reference)  \\
 (GHz)     & (mJy)        &                    \\
\hline
0.61      &   $\leq$0.9  & GMRT, 2013--2014 \citep{Kijak2017} \\
0.89      &   $\leq$1.3  & ASKAP, 2019--2020 \citep{McConnell2020} \\
1.40      &  $2.1\pm0.2$ & VLA, 2001--2004 \citep{Becker2010}  \\
3.00      & $13.6\pm1.4$ & VLA in 2019 \citep{Lacy2020}    \\
5.00      &  $5.4\pm0.3$ & VLA in 1989 \citep{Becker2010}  \\
4.86      & $26.0\pm1.3$ & VLA in 2005 \citep{Becker2010}  \\
5.00      & $23.8\pm1.3$ & WSRT on 2010 Dec 15    \\
5.00      & $24.5\pm2.0$ & EVN on 2010 Dec 15     \\
4.99      & $29.2\pm2.0$ & EVN on 2012 Feb 08     \\
4.99      & $26.4\pm1.6$ & EVN on 2012 Sep 17     \\
8.44      & $30.4\pm1.5$ & VLA on 2017 June 2     \\
9.34      & $31.9\pm1.6$ & VLA on 2017 June 2     \\
10.14     & $32.7\pm1.7$ & VLA on 2017 June 2     \\
11.04     & $33.3\pm1.7$ & VLA on 2017 June 2     \\
\hline
\end{tabular}
\end{table}

The non-simultaneous radio spectrum of \target{} is displayed in Fig.~\ref{fig:g23_alpha}. The red point marks the earliest flux density measurement observed at 5.0~GHz in 1989 \citep{Becker2010}. The blue points represent the flux densities observed between 2001 and 2019. The error bar is at the level 1$\sigma$. The arrows represent the upper limits. These flux density data are also listed in Table~\ref{tab:g23flux}. The non-simultaneous radio spectrum between 2001 and 2019 looks relatively smooth. The non-simultaneous spectral index between 1.4 and 5.0~GHz is $\sim$2.0. This is a very sharp increase and consistent with the non-detections at $<$1.4~GHz. The radio spectrum observed at $\geq$5~GHz between 2005 and 2017 follows the rising tendency, while starts to flatten significantly. It shows a flat spectrum with $\alpha = 0.34\pm0.04$ between 8 and 12~GHz.   

\target{} is an intrinsically compact source with an angular size of $\theta_{\rm int} \ll \theta_{\rm maj}$. If we use 10~mas (0.25$\theta_{\rm maj}$ at 5~GHz) as the upper limit for $\theta_{\rm int}$, its intrinsic brightness temperature is expected be at least one order of magnitude higher than the observed value, $(1.1\pm0.1) \times 10^6$~K at 5~GHz. Because of the high intrinsic brightness temperature $\geq10^{7}$~K, we identify it as a non-thermal radio source. Among all kinds of Galactic non-thermal radio sources, we cannot interpret \target{} as a pulsar wind nebula, a supernova remnant \citep[e.g.][]{Weiler2002} or a wind collision region \citep[e.g. Apep,][]{Marcote2021} in a binary. These Galactic sources usually have counterparts known from multi-wavelength observations and show a relatively extended morphology with an optically thin radio spectrum. The object \target{} is intrinsically compact and relatively bright. Its radio spectrum is optically thick at $\leq5$~GHz and flat between 8.4 and 11.0~GHz. Based on the spectrum of the source and its relatively high flux densities observed at $\geq$5~GHz for more than three decades, we cannot identify \target{} as a radio pulsar \citep[e.g.][]{Zhao2020}, a magnetar \citep[e.g.][]{Kaspi2017}, or a stellar-mass black hole in an X-ray binary system \citep[e.g.][]{Pietka2015}. Moreover, there are no X-ray, optical and infrared counterparts found by \citet{Becker2010} and by us in the latest catalogues of VizieR\footnote{\url{http://vizier.u-strasbg.fr/index.gml}} \citep{Ochsenbein2000} and the online multi-band images provided by the CORNISH survey \citep{Hoare2012}. 

\target{} is most likely an extragalactic source with a peaked radio spectrum. The rising radio spectrum of \target{} in Fig~\ref{fig:g23_alpha} is frequently seen in AGN \citep[c.f. a review by][]{ODea2021}. As a non-thermal radio source, \target{} is expected to have an optically thin radio spectrum at frequencies $>$11 GHz, and thus can be classified as a high-frequency peaked-spectrum source \citep[e.g.][]{Orienti2006, ODea2021}. Between the intrinsic peak frequency $\nu_{\rm pk}$ and the maximum linear size $l_{\rm max}$, this class of sources tend to follow a correlation, $\nu_{\rm pk} \propto l_{\rm max}^{-0.65}$ \citep{ODea1998}. According to the correlation, \target{} would have a linear size of $\la$10~pc.     

We favour an intrinsic origin for its flux density variability. At 5~GHz, the source displayed a factor of five increase of the flux density between 1990 and 2005, and then entered a relatively stable phase between 2005 and 2017. The early brightening could be driven by a sudden change of the accretion rate, a newborn jet or a change of the jet direction \citep[e.g., the extreme case NGC~660][]{Argo2015}. Generally, such large variability on timescales of decades is inconsistent with an extrinsic variability on short timescales of $\la$1 yr due to plasma propagation effects \citep[e.g.][]{Nyland2020}. Similar extragalactic variable radio sources with the convex radio spectra peaking at $\ga$3~GHz have been reported by \citet{Orienti2020} and \citet{Wolowska2021}. Their optical counterparts are most likely quasars. VLBI observations of these radio sources often reveal compact structures on the parsec scales \citep[e.g.][]{Orienti2020, Wolowska2021}. Their multi-band radio spectra also show significant variation. Some sources flattened their spectra and changed their flux densities more randomly, indicating they are significantly beamed sources and similar to blazars \citep[e.g.]{Wolowska2021}. Future multi-epoch high-frequency VLBI observations would provide key clues for us to answer whether \target{} is a beamed source and have a relativistic ejecta. Moreover, the multi-frequency monitoring of its radio spectrum also allows us to investigate whether it follows the adiabatic expansion \citep[e.g.][]{Orienti2020}.

\subsection{Contribution from the angular broadening by scattering}
\label{subsec:discussion_scatter}

All sources in the Galactic plane suffer scatter broadening to some degree. A model of electron distribution in the Galaxy presented by \citet{Cordes1991} offered a prediction of the scatter broadening in the absence of strong clumps of turbulence (see Fig.~3 in the cited paper, $\theta_{\rm scat}$ is 125--1000~mas at 1~GHz). Scaling this prediction to our observing frequency of 5~GHz, we estimate the scattering diameter $\theta_{\rm scat}$ at 5 GHz in the sky areas of our interest being in the range 5--40~mas. Our calibrator J1846$-$0651 (Galactic latitude: 0.87~deg) has an angular size of $5.03\pm0.02$~mas. Because of the very low Galactic latitude, \target{} suffer from the stronger angular broadening \citep{Cordes1991} in the sample and thus have much larger angular sizes (25--45~mas). Our results are generally consistent with the model predictions. We also notice that the scatter broadening might be very sky position dependent \citep[e.g.][]{Cordes2002, Cordes2003} and there are very limited input pulsar data at the distance of $\ga$5~kpc to constrain electron distribution models \citep[e.g.][]{Yao2017}. The available AGN size measurements are also very rare in particular at the very low Galactic latitude \citep[e.g.][]{Cordes2003, Pushkarev2015}. Thus, the model-estimated $\theta_{\rm scat}$ has a very large uncertainty (up to $\sim$1~dex). Our constraint $\theta_\textsc{fwhm}$ in Table~\ref{tab:size} might include a strong contribution from $\theta_{\rm scat}$. There might be more sources intrinsically compact on the centi-arcsec scales. 

Several sources might be slightly resolved on the arcsec scales. In Table~\ref{tab:size}, six sources (No. 2, 5, 10, 12, 15 and 17) have a size ($\theta_\textsc{vla5}$) larger than the beam size (1.5~arcsec) at the level $>$2$\sigma$ in the CORNISH images \citep{Hoare2012, Purcell2013}. Their angular sizes may not be explained as a consequence of extreme strong scatter broadening, $\theta_{\rm scat} = 0.34$~arcsec at 5~GHz \citep{Moran1990}. Moreover, two sources show a clearly-seen extension. The related VLA survey images are displayed in Fig.~\ref{fig:g37} and~\ref{fig:g22}.

\subsection{Classifications of the 17 variable sources}
\label{subsec:discussion_alpha}
Three sources can be firmly identified as Galactic objects in our sample. These Galactic objects have clearly-seen infrared counterparts in the online multi-wavelength image database collected by the survey CORNISH \citep{Hoare2012, Purcell2013}. 

Two arcsec-scale extended sources G25.7156$+$0.0488 and G37.7347$-$0.1126 are young H\,\textsc{ii} regions \citep{Becker2010, delaFuente2020, Yang2021HII}. In Table~\ref{tab:flux}, the two thermal sources, powered by free-free radiation, have flux densities $>$10 mJy at 1.06~GHz, while drops to $<$1.3~mJy at frequency 0.88~GHz. This sharp change is mainly because of free-free absorption \citep{delaFuente2020, Yang2021}.

\begin{figure}
\includegraphics[width=\columnwidth]{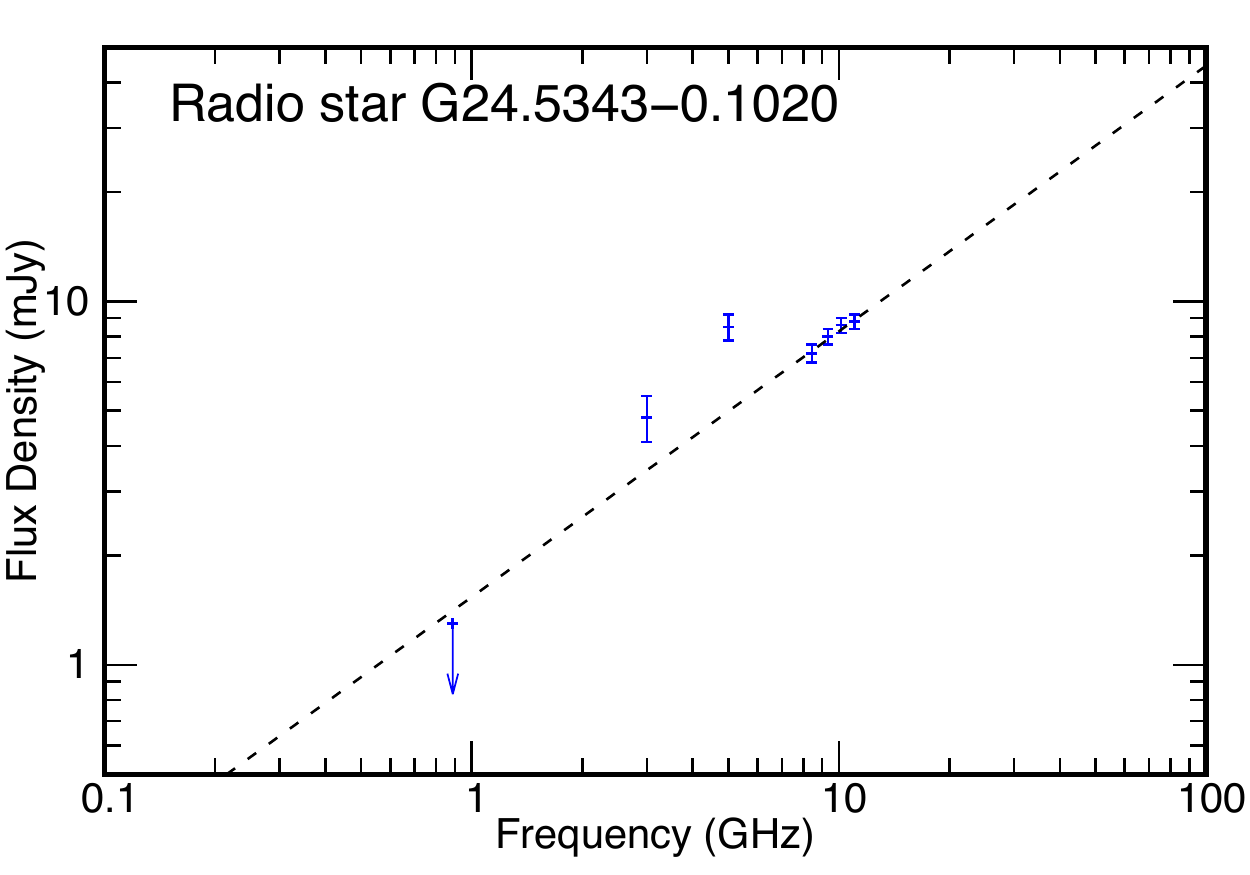}
\caption{The rising radio spectra of the variable radio star G24.5343$-$0.1020. The data are from various non-simultaneous observations between 2011 and 2020 and listed in Tables~\ref{tab:size}, \ref{tab:flux} and \ref{tab:vla017a-070}. The dashed line represents the best-fitting model $S(\nu)=(1.53\pm0.48)\nu^{0.73\pm0.14}$ derived from the data points at frequencies $>$8~GHz. }
\label{fig:g24}
\end{figure}

The source G24.5343$-$0.1020 was identified as a candidate radio star in the CORNISH catalogue \citep{Hoare2012, Purcell2013}. \citet{Marocco2021} used a program \textsc{CatWISE} to catalog sources that are selected from all-sky survey data of Wide-field Infrared Survey Explorer (\textit{WISE}) at 3.4 and 4.6~$\mu$m, and detected a proper motion, $\mu_{\rm ra} = -50.75 \pm 5.5$~mas\,yr$^{-1}$, $\mu_{\rm dec} = -48.9 \pm 9.0$ mas\,yr$^{-1}$, for the source. This input allows us to confirm G24.5343$-$0.1020 as a radio star. The radio star had a quite large variability. It had a peak brightness of $\leq0.9$ mJy at 5~GHz in 1990, while brightened significantly and reached $6.8 \pm 0.3$~mJy in 2006 \citep{Becker2010} and $8.5 \pm 0.7$~mJy in 2011. Its non-simultaneous radio spectrum is displayed in Fig.~\ref{fig:g24}. It shows an inverted radio spectrum. If the radio spectrum is because of the significant absorption at the lower frequencies, future radio observations would reveal a flat spectrum for a thermal source or a steep spectrum for a non-thermal source at the frequencies $>$12~GHz. 

The source G25.2048$+$0.1251 is very likely a compact planetary nebula according to its infrared and radio properties. Its non-simultaneous radio spectrum is displayed in Fig.~\ref{fig:g25}. This is a frequently-seen radio spectrum in planetary nebulae \citep[e.g.][]{Irabor2018}. Because of significant variability or absorption at the low frequencies, its radio spectrum cannot be accurately described by a simple power-law model.  It has a faint infrared counterpart. In the Spitzer Galactic Legacy Infrared Mid-Plane Survey Extraordinaire (GLIMPSE), it has an increasing flux density from $\sim$1.4~mJy at 3.6~$\mu$m to $\sim$19.4~mJy at 8.0~$\mu$m \citep{Benjamin2003,Churchwell2009}. It is also detected by the \textit{WISE} at 24~$\mu$m and show a compact morphology ($\la$1~arcsec) with a flux density of $\sim$9.2~mJy. Most likely, its spectral energy distribution peaks at $\la$24~$\mu$m, strongly support the classification of a planetary nebula instead of a young H\,\textsc{ii} region \citep{Irabor2018}. Moreover, some planetary nebulae are known to show large radio variability probably resulting from non-thermal radio emission mechanism \citep[e.g.][]{Cerrigone2017}. However, the VLBI non-detection at 5~GHz does not support the existence of non-thermal emission of $T_{\rm b} \geq 2 \times 10^5$~K. If it is indeed a planetary nebula, its nearly flat spectrum indicates that the dominant radio continuum emission is thermal emission at $\ga$5~GHz.   

The above candidate planetary nebula G025.2048$+$0.1251 plus the other 13 sources are listed as candidate radio AGN in the CORNISH catalogue \citep{Hoare2012, Purcell2013} mainly because of no clearly-seen counterparts in the optical and infrared wavelengths. We also searched for multi-wavelength counterparts in the catalogues of VizieR, and possible associated astrophysical masers in the online maser database \footnote{\url{http://maserdb.net/}} presented by \citet{Ladeyschikov2019} from the available literature and private catalogues. Except for G25.2048$+$0.1251, we didn't find any counterparts or associated masers in these catalogues and online multi-band image databases. For a Galactic source with an angular size in our observed range ($\la$2000~mas) at a typical distance of 10~kpc, it will have a physical size of $\la$0.1~pc and thus can vary rapidly on timescales of $\la$0.3~yr \citep[e.g.][]{Zhao2020}. However, nine sources did not vary significantly over a 4-yr timescale (c.f. Table~\ref{tab:size}). Thus, the explanation of the extragalactic radio AGN are also more consistent with our results.

\begin{figure}
\includegraphics[width=\columnwidth]{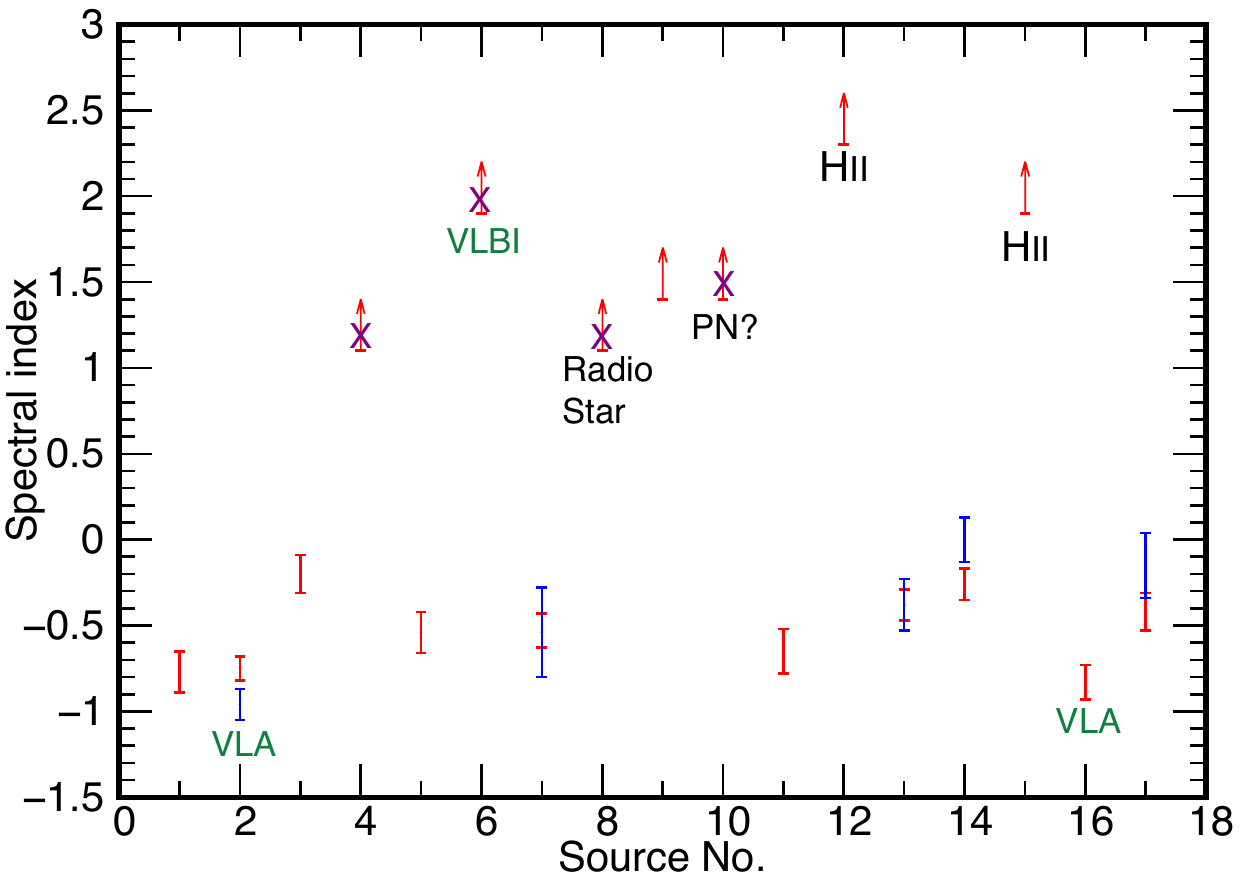} \\
\caption{
The spectral indices and the diverse nature of the sample. Red points represent the spectral indices between 0.88 and 3.00 GHz. Red arrows represent the lower limits. Blue points represent the spectral indices between 1.00 and 2.00 GHz. The error bars are at the level of 1$\sigma$. The input data are listed in Table~\ref{tab:flux}. Purple crosses show the sources observed by the Jansky VLA at X band and listed in Table~\ref{tab:vla017a-070}.  The notes in black highlight three Galactic sources and one candidate planetary nebula (PN). The notes in green indicate the sources displaying the significantly extended structures in the VLBI (cf. Fig.~\ref{fig:g23}) and VLA images (cf. Fig.~\ref{fig:g37} and \ref{fig:g22}). }
\label{fig:alpha17}
\end{figure}

\begin{figure}
\centering
\includegraphics[width=0.48\textwidth]{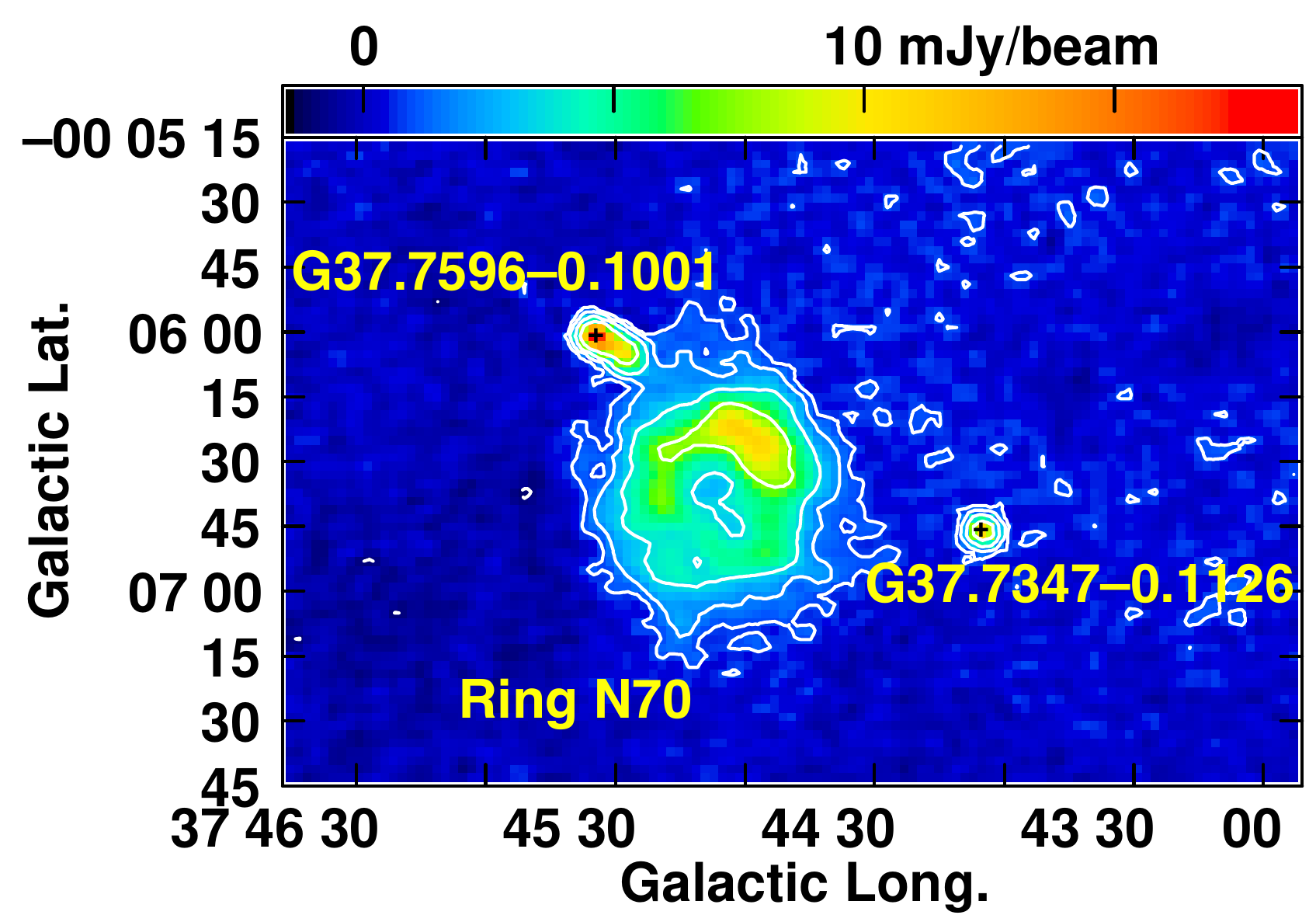}  \\
\caption{
The 1.4-GHz VLA image of the radio source G37.7596$-$0.1001 and the young H~\textsc{ii} region G37.7347$-$0.1126. It is clearly seen that G37.7596$-$0.1001 shows a 12.5-arcsec-long extension toward south. The image is from the survey MAGPIS \citep{Helfand2006}. The ring structure N70 \citep{Churchwell2006} is also associated with a Galactic H~\textsc{ii} region, G37.754$-$0.108 \citep{White2005}. The sizes of the beam are FWHM~=~$6.4 \times 5.4$~arcsec. The map has a peak brightness of 18.5~mJy~beam$^{-1}$ and a noise level of $\sigma = 0.36$~mJy~beam$^{-1}$. The contours represent the levels 2.5$\sigma$~$\times$~($-$1, 1, 2, 4, 8).  }
\label{fig:g37}
\end{figure}

\begin{figure*}
\centering
\includegraphics[height=\columnwidth]{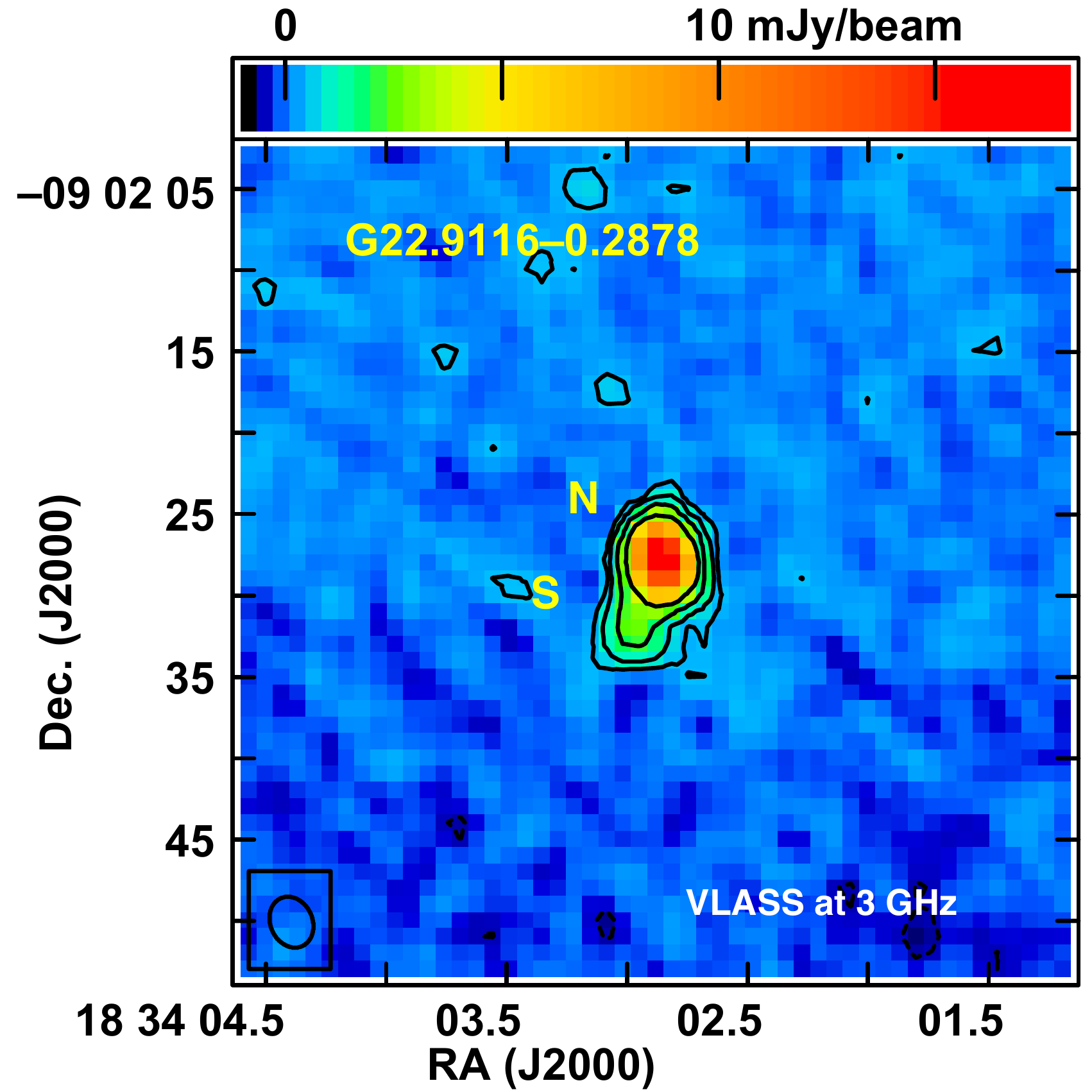}  
\includegraphics[height=1.02\columnwidth]{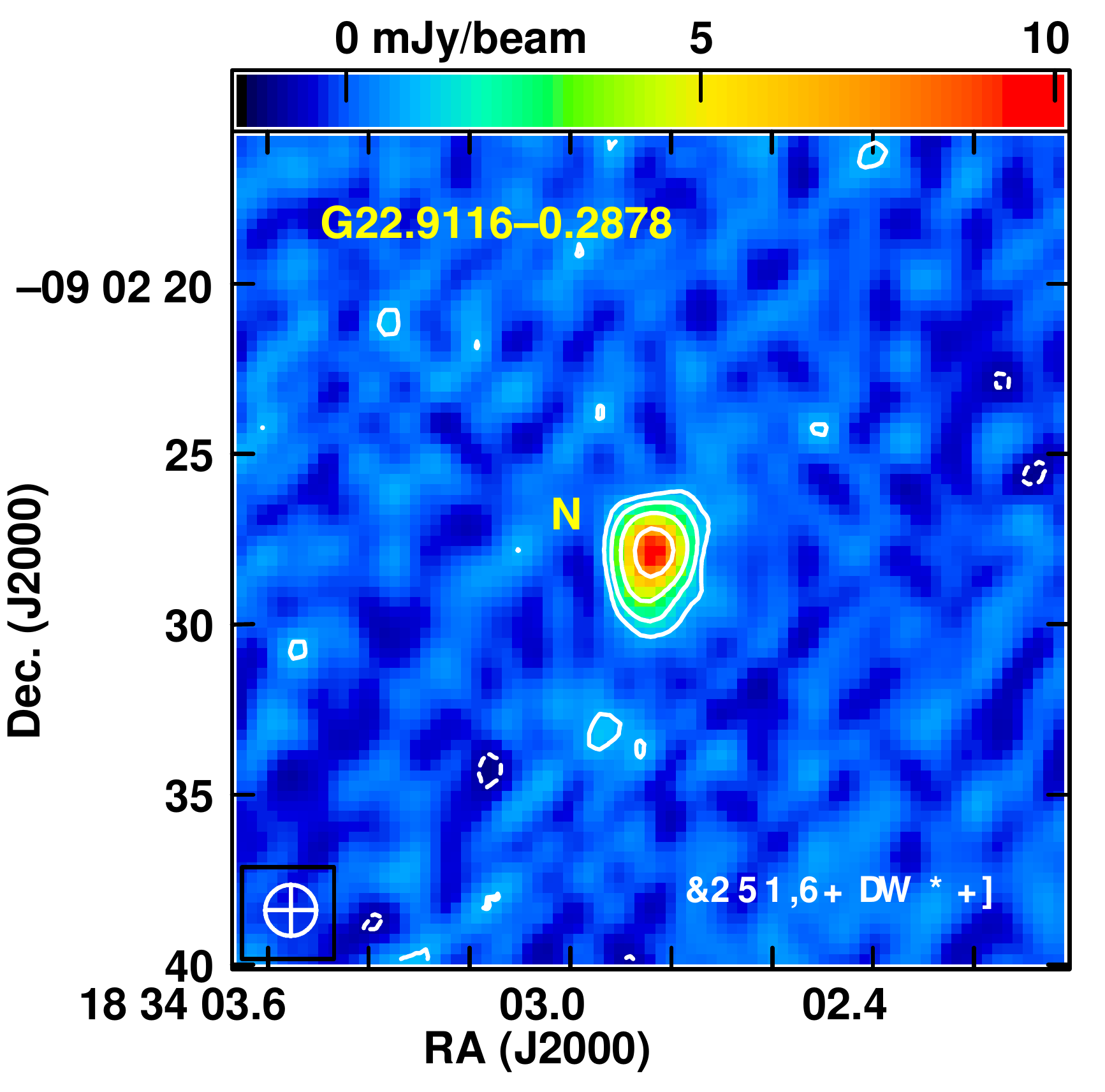} \\
\caption{
The arcsec-scale structure of the steep-spectrum source G22.9116$-$0.1939 observed by the VLA surveys VLASS \citep{Lacy2020} and CORNISH \citep{Hoare2012, Purcell2013}. The contours represent the levels 2.5$\sigma$~$\times$~($-$1, 1, 2, 4, 8). The 3-GHz map has a peak brightness of 17.8~mJy~beam$^{-1}$ and an off-source noise level of $\sigma = 0.2$~mJy~beam$^{-1}$. The 5-GHz map has a peak brightness of 10.0~mJy~beam$^{-1}$ and a noise level of $\sigma = 0.35$~mJy~beam$^{-1}$. The beam FWHM is $3.3 \times 2.2$~arcsec at 3~GHz and $1.5 \times 1.5$~arcsec at 5 GHz. The integrated flux density is $28.7 \pm 1.4$ mJy at 3~GHz and $15.5 \pm 1.6$~mJy at 5~GHz. }
\label{fig:g22}
\end{figure*}

\begin{figure}
\includegraphics[width=\columnwidth]{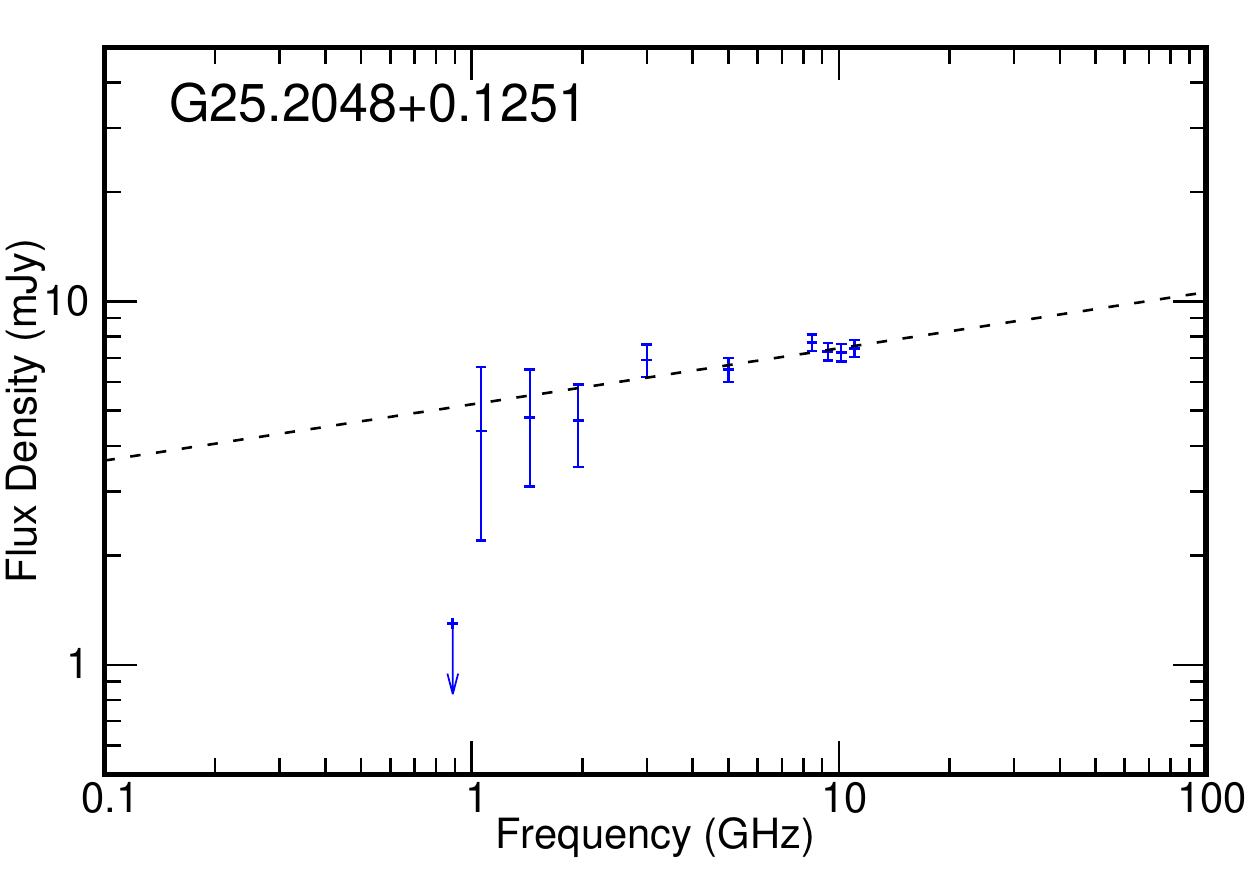} \\
\caption{The rising radio spectra of the variable source G25.2048$+$0.1251. The data are from various non-simultaneous observations between 2011 and 2020 and listed in Tables~\ref{tab:size}, \ref{tab:flux} and \ref{tab:vla017a-070}. The dashed lines represent the best-fitting model $S(\nu)=(5.2\pm0.5) \nu^{0.15\pm0.05}$ derived from the data points at frequencies $>1$~GHz.}
\label{fig:g25}
\end{figure}

The spectral indices of the 17 sources in Table~\ref{tab:flux} are also displayed in Fig.~\ref{fig:alpha17}. According to the non-simultaneous spectral indices observed by the ASKAP and the VLA over a timescale of about one year and assuming no significant ($>$15 per cent) variability, these sources can be clearly divided into two classes. Seven sources have rising spectra with spectral indices $\alpha>1$ between 0.88 and 3.00~GHz. This class of sources includes three Galactic objects, a candidate planetary nebula and the most compact source \target{} (cf. the VLBI image in Fig.~\ref{fig:g23}). The remaining ten sources have flat or steep spectra with spectral indices $\alpha \la 0$. Together with the simultaneous spectral indices between 1 and 2~GHz provided by \citet{Bihr2016}, we notice that two sources (G22.9116$-$0.2878/No. 2 in Fig.~\ref{fig:g22} and G37.7596$-$0.1101/No. 16 in Fig.~\ref{fig:g37}) have steep spectra with $\alpha < -0.5$ and one source (G26.2818$+$0.2312/No. 14) has a flat spectrum with $|\alpha| < 0.5$.    

According to the observed structures and radio spectra, three sources are very likely AGN jets. The source \target{} has been discussed in detail in Section \ref{sec:discussion_g23}. Furthermore, the steep-spectrum source G37.7596$-$0.1001 shows an elongated morphology at 1.4~GHz with a de-convolved length of $\sim$12.5~arcsec in the survey MAGPIS \citep{Helfand2006}. The structure is also shown in Fig.~\ref{fig:g37}. In the images provided by the CORNISH at 5~GHz \citep{Hoare2012, Purcell2013} and the VLASS at 3~GHz \citep{Lacy2020}, the other steep-spectrum source G22.9116$-$0.2878 shows a 4.5-arcsec extension toward south. This extension is also displayed in Fig.~\ref{fig:g22}. The faint extensions are not clearly seen in the 5-GHz CORNISH images probably because of their steep spectra. Together with their linear structure, we explains these faint arcsec-scale extensions as the collimated jet activity. 

\begin{figure}
\includegraphics[width=\columnwidth]{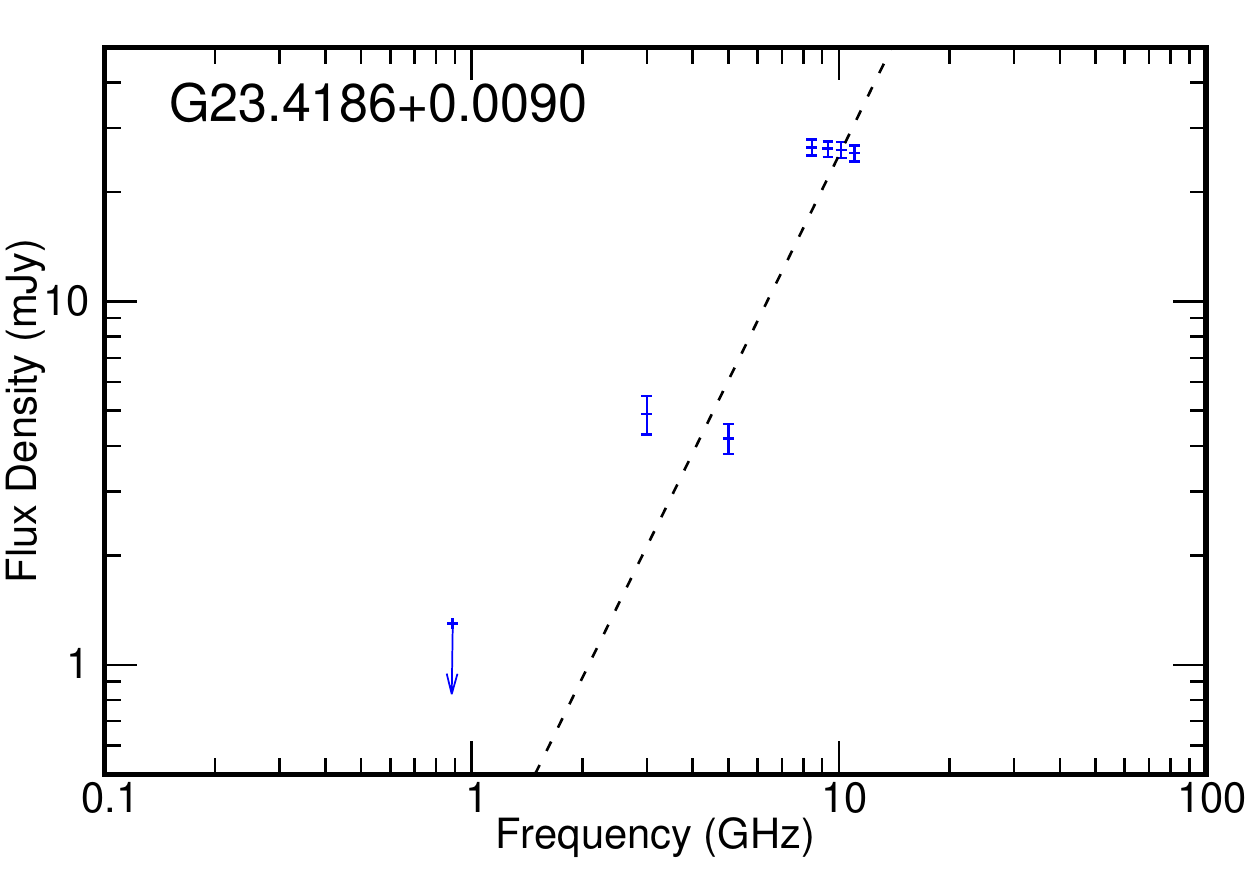} \\
\caption{The rising radio spectra of the variable source G23.4186$+$0.0090. The data are from various non-simultaneous observations between 2011 and 2020 and listed in Tables~\ref{tab:size}, \ref{tab:flux} and \ref{tab:vla017a-070}. The dashed lines represent the best-fitting model $S(\nu)=(0.22\pm0.23) \nu^{2.05\pm0.48}$ derived from the data points at frequencies $>1$~GHz.}
\label{fig:g23p4}
\end{figure} 

\begin{figure}
\includegraphics[width=\columnwidth]{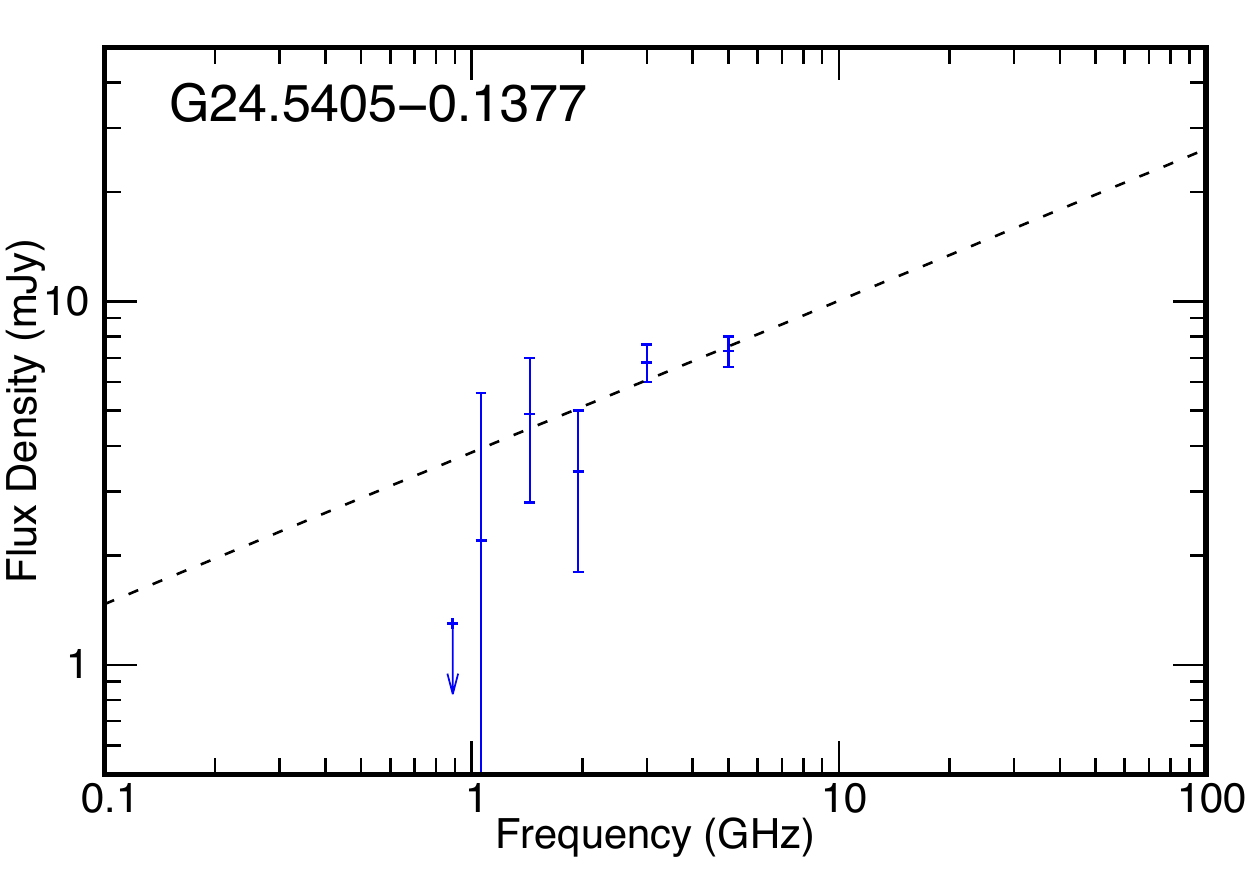} \\
\caption{The rising radio spectra of the variable source G24.5405$-$0.1377. The data are from various non-simultaneous observations between 2011 and 2020 and listed in Tables~\ref{tab:size} and \ref{tab:flux}. The dashed lines represent the best-fitting model $S(\nu)=(3.84\pm0.41) \nu^{0.42\pm0.18}$ derived from the data points at frequencies $>1$~GHz.}
\label{fig:g24p54}
\end{figure}

In the remaining eleven candidate radio AGN, three sources (No. 4, 9 \& 10) have rising radio spectra. They have a peak frequency $\ga$3~GHz. Their non-simultaneous radio spectra are displayed in Fig.~\ref{fig:g25}, \ref{fig:g23p4} and \ref{fig:g24p54}.  For each source, we tried to fit its spectrum to a simple power-law function. Because of spectral variability and possible absorption at $<$1~GHz, the model,i.e. the dashed line, give a relatively poor description to the data. They are likely intrinsically compact radio AGN with the convex radio spectra \citep{ODea2021}. Using the correlation of $\nu_{\rm pk} \propto l_{\rm max}^{-0.65}$ \citep{ODea1998}, they would have a linear size of $\la$100~pc. Their radio spectra might vary significantly and become flat at a certain time \citep[e.g.][]{Orienti2020}. Their optical counterparts are likely quasars instead of galaxies \citep[e.g.][]{Wolowska2021}.  

The other eight sources have flat or steep spectra. Compared to the peak $\alpha = -0.71$ of the spectral index distribution found by \citet{Gordon2021} in about half million radio sources between 1.4 and 3.0~GHz, these sources have relatively flatter spectra. Together with their compact morphology in the CORNISH images and their variability on decadal timescales, they would represent relatively young and compact radio AGN activity \citep[e.g.][]{ODea2021}. Some sources might be young or new-born jets on the pc scales if they have inverted radio spectra peaking at frequencies $<$0.9~GHz \citep[e.g.][]{Ross2021}. The sources G26.0526$-$0.2426,  G22.9743$-$0.3920, G26.2818$+$0.2312 and G39.1105$-$0.0160 might also be faint blazars with flat-spectrum and variable radio cores \citep[e.g.][]{Ciaramella2004, Cheng2020}. We cannot fully exclude the possibility of being bright non-thermal slow transients in their host galaxies, such as the decades-long transient FIRST~J141918.9$+$394036 \citep{Law2018}. However, this kind of possibility is very low for the small sky area $\sim$10~deg$^{2}$ \citep[e.g.][]{Murphy2021}. Furthermore, all the sources were still significantly detected in the VLASS at 3.0~GHz over about three decades since the early VLA 5-GHz Galactic plane survey between 1989 and 1991 \citep{Becker1994, Becker2010}. Their kpc-scale host galaxies and jets might show similar spectra \citep[e.g.][]{Zajacek2019} and contribute a certain diffuse radio emission to the total radio flux densities \citep[][]{Linden2020}, while their contributions are expected to be very stable on decadal timescales. Thus, we cannot associate their radio variability to their kpc-scale host galaxies or diffuse jets. Because none of these sources has an arcsec-scale extended radio lobe and a bright centi-arcsec-scale feature, we cannot explain the variability as a consequence of some hotspots in the extended lobes.

Three sources G23.5585$-$0.3241, G37.7347$-$0.1126 (H~\textsc{ii} region) and G37.7596$-$0.1001 were not detected by \citet{Becker2010} in the first epoch 1990$+$. They also cannot be identified as one-off astrophysical transients because these non-detections at 5 GHz are questionable according to the multi-frequency and multi-epoch online VLA images\footnote{\url{ https://third.ucllnl.org/cgi-bin/gpscutout}} provided by the MAGPIS \citep[][]{Helfand2006}. The source G23.5585$-$0.3241 was detected with a peak brightness of 0.75$\pm$0.15 mJy\,beam$^{-1}$ in the first epoch 1990$+$. The two sources G37.7347$-$0.1126 and G37.7596$-$0.1001 are located in a very crowded field (cf. the 1.4-GHz VLA image, Fig.~\ref{fig:g37}). The online 5-GHz VLA image of the two sources in the first epoch 1990$+$ has a poor quality and shows some strong image artefacts most likely resulting from a poor de-convolution of the complex field. Furthermore, G37.7596$-$0.1001 was detected with a peak brightness of $11.8 \pm 1.8$ mJy~beam$^{-1}$ at 1.4~GHz in 1983 \citep{Becker1990}.

According to the above discussion, we give some comments in Table~\ref{tab:size}. Our analyse shows that the sample has a divers nature. Among these variable sources of $\ga$1~mJy in the Galactic plane, AGN activity still plays a dominant role. We caution that there might still exist faint centi-arcsec-scale compact components that were missed in a certain sources because of their variability and our limited baseline sensitivity. Future multi-frequency radio observations of the sample with the electronic Multi-Element Remotely Linked Interferometer Network would have ideal resolutions to further reveal their structures and probe their natures.

\section{Conclusions}
\label{sec:conclusions}
With the EVN at 5~GHz, we observed 17 variable sources with a Galactic latitude $<$1 deg. We detected only \target{} in the high-resolution VLBI images, yet all the sources in the low-resolution WSRT images. Moreover, we performed another two-epoch EVN observations of the detected source \target{}. It had a stable flux density of $26.8 \pm 1.8$~mJy between 2010 December and 2012 September. Its radio structure is fully in agreement with an elliptical Gaussian model with the angular sizes $43.4 \pm 0.7$ mas by $28.7 \pm 1.3$~mas at position angle $39.4 \pm 2.7$~deg. Together with its inverted radio spectrum and long-term flux density variability, the structure could be interpreted as a compact extragalactic source plus a strong scatter broadening. We noticed two young H~\textsc{II} regions that were previously classified in literature, and identified a radio star and a candidate planetary nebula. Furthermore, we studied the spectral properties of the rest of the candidate radio AGN that have no optical and infrared counterparts. Using the non-simultaneous flux density measurements observed by the RACS at 0.9~GHz and the VLASS at 3.0~GHz with an interval of about one year, we found that three sources have rising spectra and the remaining ten sources have flat or steep spectra with $\alpha >-1$. In the existing VLA survey images, two sources display faint jet activity on the arcsec scales. Our study indicates that the sample has a diverse nature and AGN activity play a dominant role among variable sources of $\ga$1~mJy in the Galactic plane.

\section*{Acknowledgements}
The authors are grateful to James Miller-Jones, Pikky Atri and Arash Bahramian for a very useful discussion and permission to compare results presented here with the complementary VLBA observations before publication of the latter.   
This work is supported by the National Science Foundation of China (grants 11763002, 11590784, 11933007) and the National SKA Program of China No. 2020SKA0110100.
The EVN is a joint facility of independent European, African, Asian, and North American radio astronomy institutes. Scientific results from data presented in this publication are derived from the following EVN project code(s): EY014 and EY017. 
e-VLBI research infrastructure in Europe is supported by the European Union’s Seventh Framework Programme (FP7/2007-2013) under grant agreement number RI-261525 NEXPReS.
The research leading to these results has received funding from the European Commission Seventh Framework Programme (FP/2007-2013) under grant agreement No. 283393 (RadioNet3).
The VLA is operated by the NRAO, which is a facility of the National Science Foundation operated under cooperative agreement by Associated Universities, Inc.
This research has used NASA’s Astrophysics Data System Bibliographic Services and the NASA/IPAC Extragalactic Database (NED), which is operated by the Jet Propulsion Laboratory, California Institute of Technology, under contract with the National Aeronautics and Space Administration,
the VizieR catalogue access tool, CDS, Strasbourg, France (DOI: 10.26093/cds/vizier).
This research has made use of the CIRADA cutout service at URL cutouts.cirada.ca, operated by the Canadian Initiative for Radio Astronomy Data Analysis (CIRADA). CIRADA is funded by a grant from the Canada Foundation for Innovation 2017 Innovation Fund (Project 35999), as well as by the Provinces of Ontario, British Columbia, Alberta, Manitoba and Quebec, in collaboration with the National Research Council of Canada, the US National Radio Astronomy Observatory and Australia’s Commonwealth Scientific and Industrial Research Organisation.
\section*{Data Availability}
The correlation data of the experiments EY014, EY017A and EY017B are available in the EVN data archive (\url{http://www.jive.nl/select-experiment}). The calibrated visibility data underlying this article will be shared on reasonable request to the corresponding author. The VLA data are available in the NRAO Science Data Archive (\url{https://archive.nrao.edu/archive/advquery.jsp}. The WSRT data are available in the WSRT archive (\url{https://old.astron.nl/radio-observatory/astronomers/wsrt-archive}).



\bibliographystyle{mnras}
\bibliography{G23_MN} 







\bsp	
\label{lastpage}
\end{document}